\documentclass[final,authoryear,12pt]{elsarticle}

\usepackage{graphicx}
\usepackage{epstopdf}
\usepackage{hyperref}
\usepackage{setspace}
\usepackage{booktabs}
\usepackage{natbib}
\usepackage{bbm, dsfont}
\usepackage{amssymb}
\usepackage{amsmath,amsthm}
 \usepackage{geometry}
 \usepackage{bm}
\usepackage{color}
\usepackage{hyperref}
\usepackage{array,multirow}
\usepackage{caption}
\usepackage{lscape}
\usepackage{url}

\usepackage{color,soul}
\definecolor{lightblue}{rgb}{.80,.9,1}
\sethlcolor{lightblue}
\setulcolor{red}

\journal{Energy Journal}

\begin{document}

\begin{frontmatter}

\title{Total, asymmetric and frequency connectedness between oil and forex markets\tnoteref{label1}}

\author[ies,utia]{Jozef Barun\'{\i}k}\corref{cor2}
\author[ies]{Ev\v{z}en Ko\v{c}enda}

\address[ies]{Institute of Economic Studies, Charles University, Opletalova 26, 110 00, Prague, Czech Republic}
\address[utia]{Institute of Information Theory and Automation, The Czech Academy of Sciences, Pod Vodarenskou Vezi 4, 182 00, Prague, Czech Republic}

\cortext[cor2]{Corresponding author, Tel. +420 776 259 273, Email address: barunik@fsv.cuni.cz}

\tnotetext[label1]{We are thankful for valuable comments we received from the editor Fredj Jawadi,  V\'aclav Bro\v{z}, Julien Chevallier, Lutz Kilian, four anonymous referees, and participants in the Fifth International Symposium in Computational Economics and Finance (ISCEF) in Paris. Ko\v{c}enda acknowledges support from the GA\v{C}R grant 19-15650S. Part of the paper was written while Ko\v{c}enda was a GEMCLINE visiting researcher at the Energy Center of the University of Auckland, whose hospitality is acknowledged. The usual disclaimer applies.}

\date{no}

\begin{abstract}
\noindent 
We analyze total, asymmetric and frequency connectedness between oil and forex markets using high-frequency, intra-day data over the period 2007 -- 2017. By employing variance decompositions and their spectral representation in combination with realized semivariances to account for asymmetric and frequency connectedness, we obtain interesting results. We show that divergence in monetary policy regimes affects forex volatility spillovers but that adding oil to a forex portfolio decreases the total connectedness of the mixed portfolio. Asymmetries in connectedness are relatively small. While negative shocks dominate forex volatility connectedness, positive shocks prevail when oil and forex markets are assessed jointly. Frequency connectedness is largely driven by uncertainty shocks and to a lesser extent by liquidity shocks, which impact long-term connectedness the most and lead to its dramatic increase during periods of distress.

\end{abstract}


\end{frontmatter}


\section{Introduction}

Knowledge and quantification of the volatility connectedness, or volatility spillovers, between oil and forex markets is important because most crude oil production and sales is quoted and invoiced in US dollars \citep{devereux2010oil}, and oil prices in domestic currencies thus depend substantially on the dollar exchange rate.\footnote{In general, volatility connectedness quantifies the dynamic and directional characterization of volatility spillovers among various assets or across markets \citep{diebold2015financial}. In the text, we use the terms connectedness and spillovers interchangeably, as both have been used in the literature to describe the same phenomenon.} Payments for the oil sold on the market represent massive financial flows entering the forex market \citep{baker2018commodities}. In addition, large financial flows come from financial players with no interest in oil as a physical commodity, which contributed to the spectacular increase in the financialization of oil after 2004 and the subsequent reshaping of the oil market \citep{fattouhmahadeva2014}.\footnote{It has to be noted that \cite{Fattouh2013oilfin} find that the existing evidence is not supportive of an important role of speculation in driving the spot price of oil after 2003. Instead, there is strong evidence that the co-movement between spot and futures prices reflects common economic fundamentals rather than the financialization of oil futures markets.}  Empirical evidence also shows that oil prices possess predictive power with respect to the exchange rates of the oil-exporting countries \citep{ferraro2015can}.

In connection to the above phenomena, there is considerable potential for the uncertainty (i.e., volatility) in oil prices to transfer into the uncertainty of foreign currencies on the forex market and vice versa. As such, large volatility spillovers are likely to emerge between the two markets. The literature analyzing the nexus between oil and forex markets has investigated primarily co-movements between the two. However, to the best of our knowledge, volatility spillovers between the two markets have not been fully explored yet despite that they impact many areas of research and carry practical implications related to risk management \citep{Kanas2001}, portfolio allocation \citep{aboura2014cross}, and business cycle analysis. Hence, our goal and contribution is a comprehensive analysis of the volatility connectedness between the oil and forex markets.

We analyze three types of connectedness, and each direction is supported by specific motivation. First, total connectedness makes it possible to quantify the aggregate extent of volatility spillovers between the two markets. The oil market exhibits historically high volatility that also surpasses that of other energy commodities \citep{regnier2007oil}.  Oil price volatility is particularly important because it represents risk to producers and industrial consumers in terms of production, inventories, and transportation, and it also affects the decisions of purely financial investors \citep{pindyck2004oil} and decisions on strategic investments \citep{henriques2011effect}. The sheer extent of transactions related to oil is likely to produce substantial volatility spillovers. However, their extent might change when combined with the relevant forex transactions. Hence, this part of our analysis enables us to assess how the aggregate level of connectedness between the two markets evolves. We also relate dynamics in the total connectedness to major economic conditions and events that affect both types of studied assets. We analyze the aggregate connectedness between the oil and forex markets with the volatility spillover index (the DY index) of \cite{diebold2009measuring} that was further improved in \cite{diebold2012better}.

Second, we extend our analysis to account for potential asymmetries in connectedness. \cite{narayan2007modelling} show that negative and positive shocks produce asymmetric effects on oil price volatility. Further, \cite{kilian2009not} shows that demand shocks related to the potential future shortfalls in oil supply affect oil prices more than actual physical supply shocks do. Since oil is an asset for which spillovers have historically played a prominent role \citep{haigh2002crack}, the existence of asymmetries in oil price volatility might naturally lead to asymmetries in volatility spillovers. Subsequently, currencies on the forex market would be able to continuously absorb or transfer those asymmetries because of the 24-hour operation of the global forex market. Moreover, \cite{barunik2017} show that currencies exhibit asymmetric connectedness. These asymmetries might be transferred via the US dollar or other key currencies, as the forex market exhibits a very high degree of integration, especially for the key currencies \citep{kitamura2010testing}. We account for asymmetric sources of volatility by computing the DY index with the realized semivariances following \cite{shephard2010measuring}. \cite{barunik2016asymmetric} combined the DY index with realized semivariances and produced a flexible measure allowing for dynamic quantification of asymmetric connectedness.
Third, we assess frequency connectedness to distinguish whether connectedness is formed at shorter or longer frequencies, i.e., shorter or longer investment horizons. \cite{barunik2018measuring} argue that shocks to economic activity impact variables at various frequencies with various strengths, and to understand the sources of connectedness in an economic system, it is crucial to understand the frequency dynamics of connectedness. The key reason is that agents operate on different investment horizons -- these are associated with various types of investors, trading tools, and strategies that correspond to different trading frequencies \citep{genccay2010asymmetry,conlon2016commodity}. Shorter or longer frequencies are the result of the frequency-dependent formation of investors' preferences, as shown in the modeling strategies of \cite{bandi2017horizon,cogley2001frequency,ortu2013long}. In our analysis, we consider the long-, medium-, and short-term frequency responses to shocks and analyze financial connectedness at a desired frequency band. We compute frequency connectedness based on the approach of \cite{barunik2018measuring}.

The remainder of the paper is organized as follows. In Section \ref{sec:lit}, we provide an overview of the literature related to volatility spillovers on the oil and forex markets. In Section \ref{sec:metodology}, we formally introduce the methodological approach. The data are described in Section \ref{sec:Data}. In three separate subsections of Section \ref{sec:results}, we present our results for total, asymmetric, and frequency connectedness. Conclusions are offered in Section \ref{sec:conclusion}. Finally, supplementary material to this paper (available at \url{https://ideas.repec.org/p/arx/papers/1805.03980.html}) provides additional information that we will refer to as supplementary material in the rest of the text.

\section{Literature review \label{sec:lit}}

The growing oil market has become the world's largest commodity market, and oil trading has been transformed from a primarily physical product activity into a sophisticated financial market \citep{manera2013oilfin}. However, the research related to volatility spillovers among oil-based commodities is limited. \cite{haigh2002crack} analyze the effectiveness of crude oil, heating oil, and unleaded gasoline futures in reducing price volatility for an energy trader and show that uncertainty is reduced significantly when volatility spillovers are considered in the hedging strategy. \cite{hammoudeh2003causality} analyzed the volatility spillovers of the same three major oil commodities and showed the impact of different trading centers. \cite{lin2001spillover} found substantial spillover effects when the two major markets for crude oil (NYMEX and London's International Petroleum Exchange) are trading simultaneously. \cite{chang2010analyzing} found volatility spillovers and asymmetric effects across four major oil markets: West Texas Intermediate (USA), Brent (North Sea), Dubai/Oman (Middle East), and Tapis (Asia-Pacific). 

\cite{restrepo2018financial} analyze the spillover effect of stock markets' VIX on crude oil and document large similarities in the correlation dynamics between the crude oil and stock volatility series. \cite{barunik2015} quantify volatility spillovers among crude oil, gasoline, and heating oil and show that asymmetries in overall volatility spillovers due to negative (price) returns materialize to a greater extent than those due to positive returns. Their occurrence also frequently indicates the extent of real or potential crude oil unavailability, which is in line with the arguments of \cite{kilian2009not}.

Analyses of forex volatility spillovers based on the DY index remain rare. \cite[Chapter 6]{diebold2015financial} analyze the exchange rates of nine major currencies with respect to the U.S. dollar from 1999 until mid-2013. They show that forex market connectedness increased only mildly after the 2007 financial crisis and that the euro/U.S. dollar exchange rate exhibits the highest volatility connectedness among all analyzed currencies. \cite{greenwoodrisk} generalize the connectedness framework and analyze risk-return spillovers among the G10 currencies between 1999 and 2014. They find that spillover intensity is countercyclical and that volatility spillovers across currencies increase during crisis periods. Similarly, \cite{bubak2011volatility} document statistically significant intra-regional volatility spillovers among the European emerging forex markets and show that volatility spillovers tend to increase in periods characterized by high market uncertainty. Further, \cite{mcmillan2010return} and \cite{antonakakis2012exchange} document the existence of volatility spillovers among the exchange rates of major currencies. Finally, \cite{barunik2017} document sizable asymmetries in volatility spillovers among the most actively traded currencies.

The recent connection between the oil and forex markets was identified by \cite{aloui2013conditional}, who analyze the dependence structure between crude-oil spot prices (WTI Cushing and Brent price indices) and nominal exchange rates of the U.S. dollar against five major currencies (euro, Canadian dollar, British pound, Swiss franc, and Japanese yen). Their results reveal the existence of a dependence structure between the two markets over the 2000 -- 2011 period along with a significant and symmetric dependence for almost all analyzed oil-exchange rate pairs. An increase in the price of oil is found to be associated with the depreciation of the dollar. The results resonate well with the earlier findings shown for the period before the global financial crisis \citep{akram2009commodity,narayan2008understanding}.

The first strand of the above studies, broadly speaking, assesses volatility spillovers separately on either the oil or forex market. The second strand analyzes co-movements between the oil and forex markets. However, we must stress that the reviewed studies do not assess the volatility connectedness (spillovers) between the two markets. In effect, and to the best of our knowledge, an analysis of the connectedness between oil and forex markets is missing in the literature.\footnote{We acknowledge that volatility spillovers between oil and stock prices were analyzed by  \cite{arouri2012impacts} and \cite{antonakakis2018oil}, including optimal portfolio allocation between oil and stocks \citep{antonakakis2019oilimpliedvolatilities}.} In our paper, we contribute to the literature by pursuing such an analysis.

\section{Measuring total, asymmetric, and frequency spillovers\label{sec:metodology}}

The spillovers measures introduced by \cite{diebold2009measuring} are based on variance decomposition from vector autoregressions (VARs) that traces how much of the future error variance of a variable $j$ is due to innovations in another variable $k$. For $N$ assets, we consider an $N$-dimensional vector of realized volatilities, $\mathbf{RV_t} = (RV_{1t},\ldots,RV_{Nt})'$, to measure total volatility spillovers. In addition, to measure asymmetric volatility spillovers, we decompose daily volatility into negative (and positive) semivariances $\mathbf{RS_t^{-}} = (RS^{-}_{1t},\ldots,RS^{-}_{Nt})'$ and $\mathbf{RS_t^{+}} = (RS^{+}_{1t},\ldots,RS^{+}_{Nt})'$ that provide proxies for downside and upside risk. Using semivariances allows us to measure the spillovers from bad and good volatility and test whether they are transmitted in the same magnitude \citep{barunik2016asymmetric}. Details on the computation of realized semivariances are described in the supplementary material.

Let us model the $N$-dimensional vector $\mathbf{RV_t}$ by a weakly stationary VAR ($p$) as $\mathbf{RV_t} = \sum_{\ell=1}^p \mathbf{\Phi}_\ell \mathbf{RV}_{t-\ell}+ \boldsymbol{\epsilon}_t$, where $\boldsymbol{\epsilon}_t\sim N(0,\mathbf{\Sigma}_{\epsilon})$ is a vector of $iid$ disturbances, and $\mathbf{\Phi}_\ell$ denotes $p$ coefficient matrices. For the invertible VAR process, the moving average representation has the following form:
\begin{equation}
\mathbf{RV}_t = \sum_{\ell=0}^{\infty}\mathbf{\Psi}_{\ell}\boldsymbol{\epsilon}_{t-\ell}.
\end{equation}
The $N\times N$ matrices holding coefficients $\mathbf{\Psi}_\ell$ are obtained from the recursion $\mathbf{\Psi}_\ell = \sum_{j=1}^p\mathbf{\Phi}_j \mathbf{\Psi}_{\ell-j}$, where $\mathbf{\Psi}_0=\mathbf{I}_N$ and $\mathbf{\Psi}_\ell = 0$ for $\ell<0$. The moving average representation is convenient for describing the VAR system's dynamics since it allows us to isolate the forecast errors further used for computation of the connectedness of the system. \cite{diebold2012better} further use the generalized VAR of \cite{koop1996impulse} and \cite{pesaran1998generalized} to obtain forecast error variance decompositions that are invariant to variable ordering in the VAR model, and it also explicitly accommodates the possibility of measuring directional volatility spillovers.\footnote{The generalized VAR allows for correlated shocks; hence, the shocks to each variable are not orthogonalized.}

\subsection{Total spillovers\label{sec:tot}}
In order to define the total spillovers index of \cite{diebold2012better}, we consider the $H$-step-ahead generalized forecast error variance decomposition matrix having the following elements for $H=1,2,\ldots$.
\begin{equation}
\theta_{jk}^H=\frac{\sigma_{kk}^{-1}\sum_{h=0}^{H-1}\left( \mathbf{e}'_j \mathbf{\Psi}_h \mathbf{\Sigma}_{\epsilon}\mathbf{e}_k \right)^2}{\sum_{h=0}^{H-1}\left( \mathbf{e}'_j \mathbf{\Psi}_h \mathbf{\Sigma}_{\epsilon}\mathbf{\Psi}'_h\mathbf{e}_k \right)}, \hspace{10mm} j,k=1,\ldots, N,
\end{equation}
where $\mathbf{\Psi}_h$ are moving average coefficients from the forecast at time $t$; $\mathbf{\Sigma}_{\epsilon}$ denotes the variance matrix for the error vector, $\boldsymbol{\epsilon}_t$; $\sigma_{kk}$ is the $k$th diagonal element of $\mathbf{\Sigma}_{\epsilon}$; and $\mathbf{e}_j$ and $\mathbf{e}_k$ are the selection vectors, with one as the $j$th or $k$th element and zero otherwise. Normalizing elements by the row sum as $\widetilde{\theta}_{jk}^H = \theta_{jk}^H / \sum_{k=1}^N \theta_{jk}^H$, \cite{diebold2012better} then define the total connectedness as the contribution of connectedness from volatility shocks across variables in the system to the total forecast error variance:
\begin{equation}
\label{stot}
\mathcal{S}^H=100\times \frac{1}{N} \sum_{\substack{j,k=1\\ j\ne k}}^N\widetilde{\theta}_{jk}^H.
\end{equation}
Note that $\sum_{k=1}^N \widetilde{\theta}_{jk}^H=1$ and $\sum_{j,k=1}^N \widetilde{\theta}_{jk}^H=N$. Hence, the contributions of connectedness from volatility shocks are normalized by the total forecast error variance. To capture the spillover dynamics, we use a 200-day rolling window running from point $t-199$ to point $t$. Further, we set a forecast horizon $H=10$ and a VAR lag length of 2 based on the AIC.\footnote{In addition, we provide sensitivity analysis with different window lengths, horizons, and lags in the VAR system in the supplementary material.}

\subsection{Directional spillovers \label{sec:dir}}
The total connectedness indicates how shocks to volatility spill over throughout the system. However, it is also interesting to identify how individual elements of the system influence the overall system, as well as how the system influences the individual elements. Following \cite{diebold2012better}, we measure the directional spillovers received by asset $j$ from all other assets $k$ as $\mathcal{S}_{N,j\leftarrow\bullet}^H=100\times \frac{1}{N} \sum_{\substack{k=1\\ j\ne k}}^N\widetilde{\theta}_{jk}^H$, i.e., we sum all numbers in rows $j$, except the terms on a diagonal that correspond to the impact of asset $j$ on itself. The $N$ in the subscript denotes the use of an $N$-dimensional VAR. Conversely, the directional spillovers transmitted by asset $j$ to all other assets $k$ can be measured as the sum of the numbers in the column for the specific asset, again except the diagonal term $\mathcal{S}_{N,j\rightarrow\bullet }^H=100\times \frac{1}{N} \sum_{\substack{k=1\\ j\ne k}}^N\widetilde{\theta}_{kj}^H.$

\subsection{Measuring asymmetric spillovers \label{sec:MAS}}

Being able to account for spillovers from volatility due to negative returns ($\mathcal{S}^-$) and positive returns ($\mathcal{S}^+$) using realized semivariances, as well as directional spillovers from volatility due to negative returns ($\mathcal{S}_{j\leftarrow\bullet}^-$, $\mathcal{S}_{j\rightarrow\bullet}^-$) and positive returns ($\mathcal{S}_{j\leftarrow\bullet}^+$, $\mathcal{S}_{j\rightarrow\bullet}^+$) using realized semivariances, we are able to measure how information transmission mechanism is symmetric.\footnote{For a verbal interpretation of asymmetries, we adopt the terminology established in the literature \citep{patton2014good} that distinguishes asymmetries in spillovers originating due to qualitatively different uncertainty. Hence, we label spillovers as bad or good volatility spillovers (or negative or positive spillovers). Note that we drop the $H$ index to ease the notational burden. Details on the computation of realized semivariances are described in \ref{sec:semi}.}

If the contributions of $RS^-$ and $RS^+$ are equal, the spillovers are symmetric, and we expect the spillovers to be of the same magnitude as spillovers from $RV$. On the other hand, the differences in the realized semivariances result in asymmetric spillovers. Following \cite{barunik2016asymmetric}, we use bootstrapping to test the null hypothesis $\mathcal{H}^1_0:\mathcal{S}^- = \mathcal{S}^+$ of a symmetric transmission mechanism.

\subsubsection{Spillover asymmetry measure}
\label{samsec}
In order to better quantify the extent of volatility spillovers, we introduce a spillover asymmetry measure. If the negative and positive realized semivariances contribute to the total variation of returns in the same magnitude, the spillovers from volatility due to negative returns ($\mathcal{S}^-$) and positive returns ($\mathcal{S}^+$) will be equal to the spillovers from $RV$, and the null hypothesis $\mathcal{H}^1_0:\mathcal{S}^- = \mathcal{S}^+$ would not be rejected. This motivates a definition of the spillover asymmetry measure ($\mathcal{SAM}$) simply as the difference between positive and negative spillovers:
\begin{equation}
\label{sam}
\mathcal{SAM} =  \mathcal{S}^+-\mathcal{S}^-,
\end{equation}
where $\mathcal{S}^+$ and $\mathcal{S}^-$ are volatility spillover indices due to positive and negative semivariances, $RS^+$ and $RS^-$, respectively, with an $H$-step-ahead forecast at time $t$. $\mathcal{SAM}$ defines and illustrates the extent of asymmetry in spillovers due to $RS^-$ and $RS^+$. When $\mathcal{SAM}$ takes the value of zero, spillovers coming from $RS^-$ and $RS^+$ are equal. When $\mathcal{SAM}$ is positive, spillovers coming from $RS^+$ are larger than those from $RS^-$, and the opposite is true when $\mathcal{SAM}$ is negative.

\subsection{Frequency decompositions of spillover measures} 
\label{ssub:spectra}

A natural way to describe the frequency dynamics (whether long, medium, or short term) of connectedness is to consider the spectral representation of variance decompositions based on frequency responses to shocks instead of impulse responses to shocks. As a building block, \cite{barunik2018measuring} considers a frequency response function, $\boldsymbol \Psi(e^{-i \omega})=\sum_h e^{-i\omega h} \boldsymbol \Psi_h$, which can be obtained as a Fourier transform of the coefficients $\boldsymbol \Psi_h$, with $i=\sqrt{-1}$. The spectral density of $\mathbf{RV_t}$ at frequency $\omega$ can then be conveniently defined as a Fourier transform of the MA($\infty$) filtered series as
$$
\boldsymbol S_\mathbf{RV}(\omega)=\sum_{h=-\infty}^{\infty} E(\mathbf{RV}_t\mathbf{RV}'_{t-h})e^{-i\omega h}=\boldsymbol \Psi(e^{-i \omega}) \boldsymbol \Sigma \boldsymbol \Psi'(e^{+i\omega})
$$
The power spectrum $\boldsymbol S_\mathbf{RV}(\omega)$ is a key quantity for understanding frequency dynamics, since it describes how the variance of the $\mathbf{RV}_t$ is distributed over the frequency components $\omega$. Using the spectral representation for covariance, \emph{i.e.}, $E(\mathbf{RV}_t\mathbf{RV}'_{t-h})=\int_{-\pi}^{\pi} \boldsymbol S_\mathbf{x}(\omega) e^{i \omega h }d \omega$, \cite{barunik2018measuring} naturally define the frequency domain counterparts of variance decomposition.

The spectral quantities are estimated using standard discrete Fourier transforms. The cross-spectral density on the interval $d = (a,b): a,b \in \left( -\pi, \pi \right), a < b$ is estimated as $\sum_{\omega} \widehat{\boldsymbol \Psi}(\omega) \widehat{\boldsymbol \Sigma} \widehat{\boldsymbol \Psi}'(\omega),$ for $\omega \in \left\{ \left\lfloor \frac{aH}{2 \pi} \right\rfloor, ..., \left\lfloor \frac{bH}{2 \pi} \right\rfloor \right\}$, where $\widehat{\boldsymbol \Psi}(\omega) = \sum_{h=0}^{H-1} \widehat{\boldsymbol \Psi}_h e^{-2 i \pi \omega/H},$ and $\widehat{\boldsymbol \Sigma} = \widehat{\boldsymbol \epsilon}'\widehat{\boldsymbol \epsilon}/(T-z)$, where $z$ is a correction for a loss of degrees of freedom, and it depends on the VAR specification.

The decomposition of the impulse response function at the given frequency band can be estimated as $\widehat{\boldsymbol \Psi}(d) = \sum_{\omega} \widehat{\boldsymbol \Psi}(\omega)$. Finally, the generalized variance decompositions at a desired frequency band are estimated as
$$\widehat{\boldsymbol \theta}_{j,k}(d) = \sum_{\omega} \widehat{\Gamma}_j(\omega) \frac{\widehat{\sigma}_{kk}^{-1} \left(\mathbf{e}'_j \widehat{\boldsymbol \Psi}(\omega) \widehat{\boldsymbol \Sigma} \mathbf{e}_k\right)^2}{\mathbf{e}'_j \widehat{\boldsymbol \Psi}(\omega) \widehat{\boldsymbol \Sigma} \widehat{\boldsymbol \Psi}'(\omega)\mathbf{e}_j},$$

where $\widehat{\Gamma}_j(\omega)=\frac{\mathbf{e}'_j \widehat{\boldsymbol \Psi}(\omega) \widehat{\boldsymbol \Sigma} \widehat{\boldsymbol \Psi}'(\omega)\mathbf{e}_j}{\mathbf{e}'_j \Omega \mathbf{e}_j},$ is an estimate of the weighting function, where $\Omega = \sum_{\omega} \widehat{\boldsymbol \Psi}(\omega) \widehat{\boldsymbol \Sigma} \widehat{\boldsymbol \Psi}'(\omega)$.

Then, the connectedness measures at a given frequency band of interest can be readily derived by substituting the $\widehat{\boldsymbol \theta}_{j,k}(d)$ estimate into the traditional measures outlined above.\footnote{The entire estimation is done using the package \texttt{frequencyConnectedness} in \textsf{R} software. The package is available on CRAN or at \url{https://github.com/tomaskrehlik/frequencyConnectedness}.}

\section{Data\label{sec:Data}}

In this paper, we compute volatility spillover measures on the futures contracts on (i) crude oil and (ii) foreign exchange (for six currencies) over the period from January 2, 2007 to December 31, 2017. We use 5-minute intraday prices of futures contracts that are automatically rolled over to provide continuous price records.\footnote{We employ 5-minute frequency data, as they have become an established standard in the literature \citep{andersen1997intraday,andersen2003micro}. The 5-minute data provide a good trade-off between autocorrelation and microstructure noise.} The intra-day returns are computed from log-prices. The currencies are the Australian dollar (AUD), Canadian dollar (CAD), British pound (GBP), euro (EUR), Japanese yen (JPY), and Swiss franc (CHF). All these currency contracts are quoted against the U.S. dollar, i.e., one unit of a currency in terms of the U.S. dollar. This is a typical approach in the forex literature -- any potential domestic (U.S.) shocks are integrated into all currency contracts. The currencies under investigation constitute a group of the globally most actively traded currencies and constitute two-thirds of the global forex turnover by currency pair \citep{bis2016,antonakakis2012exchange}.

The crude oil futures contracts are traded on the New York Mercantile Exchange (NYMEX), and transactions are recorded in Eastern Time (EST). The foreign exchange futures contracts are traded on the Chicago Mercantile Exchange (CME) on a nearly 24-hour basis, and transactions are recorded in Central time (CST). Trading activity begins at 5:00 pm CST and ends at 4:00 pm CST. Because of this one-hour gap in trading, we redefine the day in accordance with the electronic trading system. Furthermore, similar to \cite{andersen2003micro}, we eliminate transactions executed on U.S. federal holidays, December 24-26 and December 31-January 2, because of the low liquidity on these days, which could lead to estimation bias. The data are available from Tick Data, Inc.

\begin{figure}[ht!]
   \centering
   \includegraphics[width=4in]{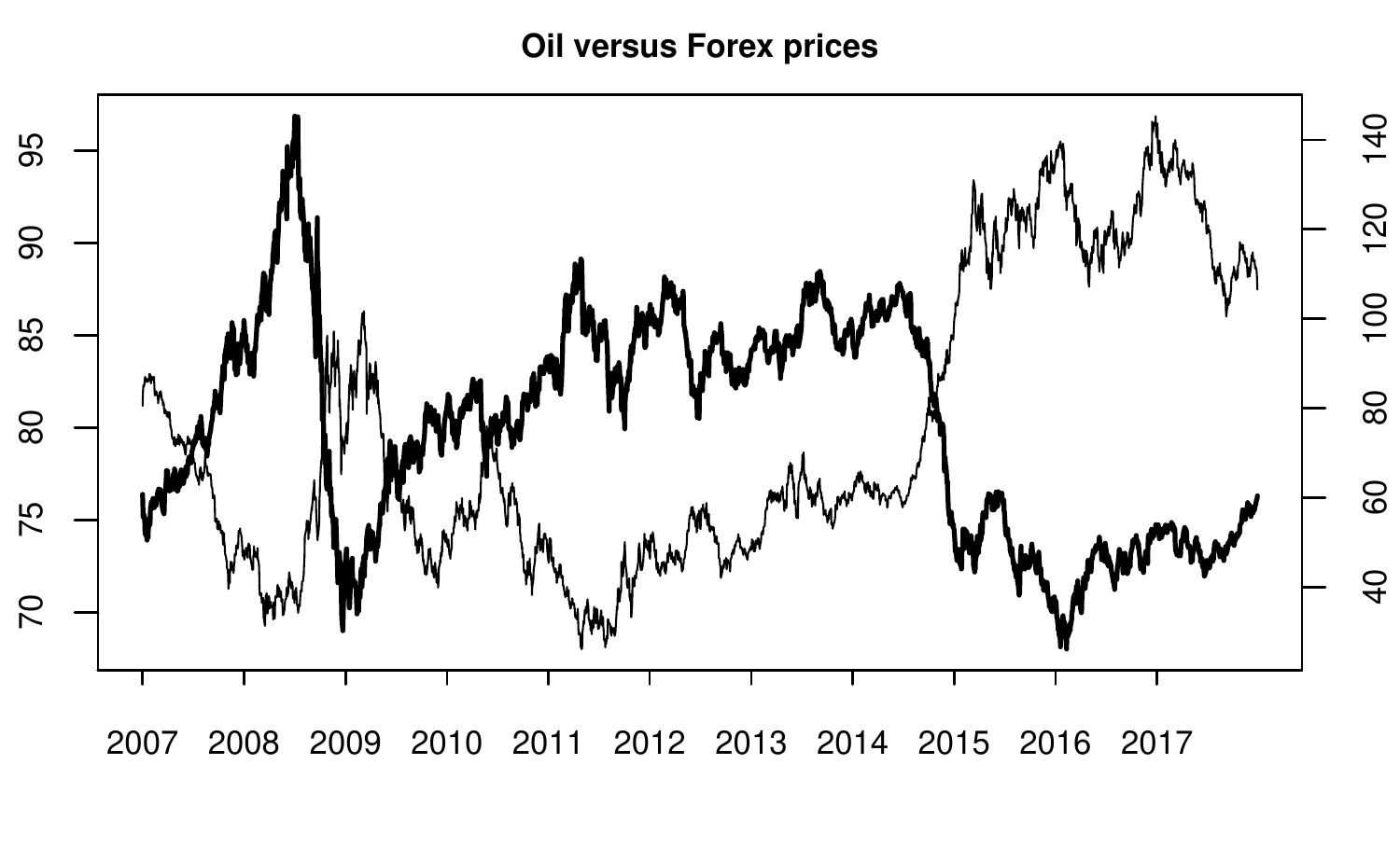}
   \caption{Trade Weighted U.S. Dollar Index (solid line and left axis), crude oil (bold solid line and right axis}
   \label{Fig5}
\end{figure}

In Figure \ref{Fig5}, we present the aggregate developments on both oil and forex markets over the period 2007 -- 2017. Instead of plotting six individual exchange rates, we use the Trade Weighted U.S. Dollar Index: Major Currencies (DTWEXM). Since the dollar index traces the dollar value against major world currencies, in a single graph, we plot the forex and crude oil prices without needing to plot developments in individual currencies. Oil prices and the dollar index fluctuate widely over the period under investigation, which is in line with earlier assessments \citep{regnier2007oil,barunik2015,barunik2017}. The common pattern reveals that a rise (drop) in the oil price is associated with depreciation (appreciation) of the U.S. dollar for most of the pictured period. This pattern is in accord with earlier findings for the pre-crisis period \citep{narayan2008understanding,akram2009commodity}. \cite{aloui2013conditional} document that oil price increases are associated with dollar depreciation for most bilateral exchange rates, including after the financial crisis.

However, the above pattern does not hold universally. In 2010 and during late 2011-2014, rising oil prices correlate with appreciation of the dollar. Heterogeneity in the pattern might be grounded in the differences across countries with respect to oil dependence. \cite{lizardo2010oil} show that rising oil prices lead to dollar depreciation with respect to currencies of countries that are net oil exporters (Canada, Mexico, and Russia) or countries that are neither net exporters nor significant importers of oil relative to their total trade (the U.K. and EU). On the other hand, an increase in oil prices leads to dollar appreciation with respect to the currencies of the net oil-importing countries such as Japan.\footnote{Heterogeneity in the pattern can also be associated with other factors, including US dependence on oil and net export position as well as the business cycle and source of a specific shock. These issues are beyond the scope of the paper and are left for further research.} Overall, Figure \ref{Fig5} provides persuasive evidence of strong linkages between the oil and forex markets.

\section{Results: Total, asymmetric and frequency connectedness \label{sec:results}}

\subsection{Total connectedness}

\begin{figure}
   \centering
   \includegraphics[width=4in]{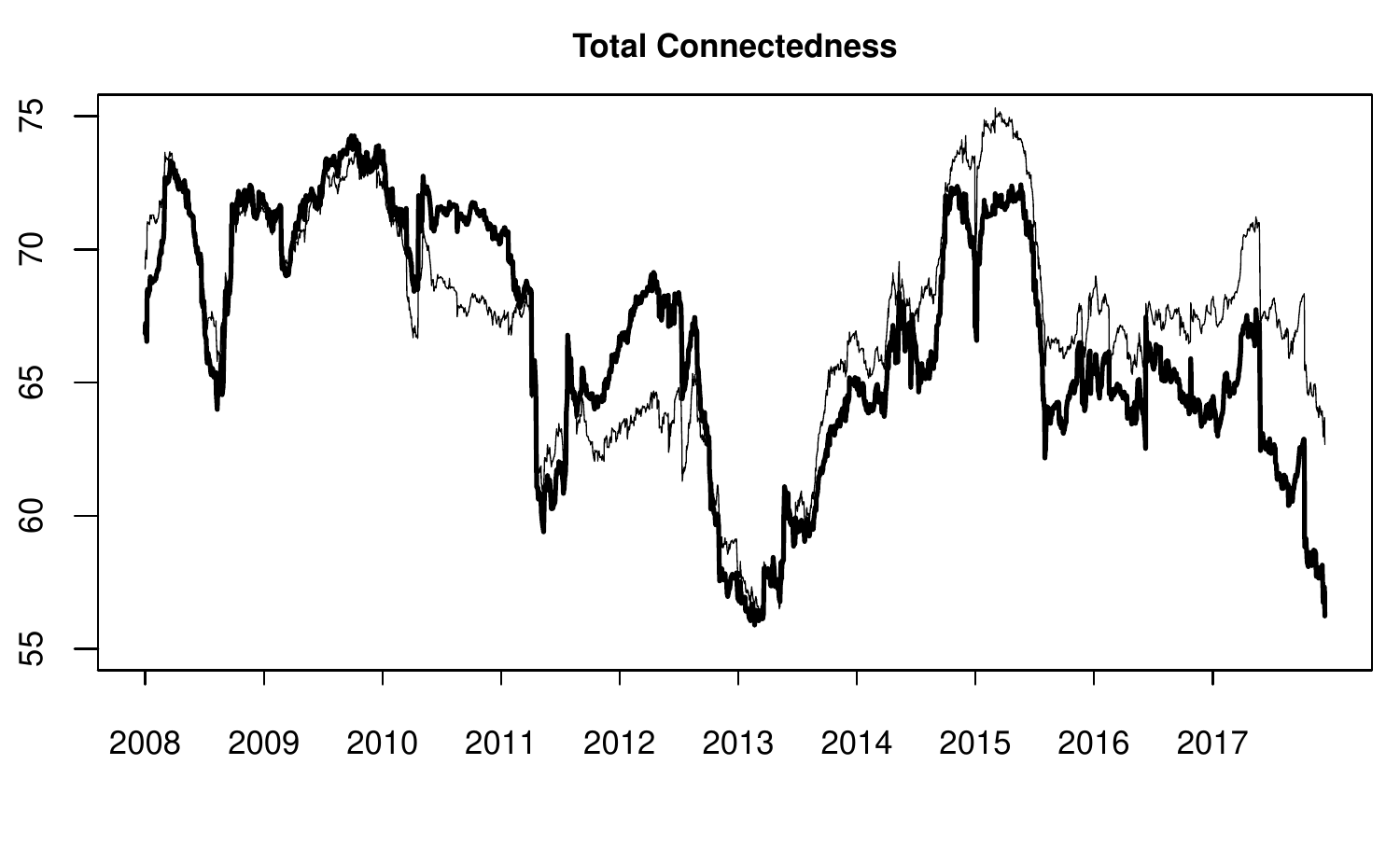}
   \caption{The total volatility connectedness of six currencies (solid line) and the total volatility spillovers of six currencies and crude oil (bold solid line).}
   \label{Fig2}
\end{figure}

In Figure \ref{Fig2}, we present the total connectedness among the six currencies (solid line) along with the total connectedness among the currencies and oil (solid bold line). The total volatility spillovers measure is calculated based on \cite{diebold2012better}. First, we examine the connectedness on the forex market (solid line): the connectedness is quite high during the GFC period until 2010 and then in 2012 and early 2014. The total connectedness values of 65\% and above during the 2008 -- 2010 period are comparable to those found in \cite[Chapter 5]{diebold2015financial}. The plot exhibits a visible structural change in total connectedness among the six currencies under investigation: an initial high connectedness is interrupted by a short drop during 2009 and decreases gradually after 2010, but then in 2013, it begins to rise. The period is marked by two distinctive phenomena. One is the difference between monetary policies among the Fed, ECB, and Bank of Japan. While the Fed stopped its quantitative easing (QE) policy in 2014, the ECB was beginning to pursue one, and the Bank of Japan was already active in pursuing this policy. From 2013, the policy differences affected the capital flows and carry-trade operations such that the U.S. dollar began to appreciate against the euro and yen. The two simultaneously falling commodity prices exerted downward pressure on inflation and interest rates. This course affects most currencies in our sample, as commodities are quoted in vehicle currencies (USD, EUR, JPY), and interest rate cuts occurred for commodity currencies (AUD, CAD), diminishing their appeal for carry-trade activities.

The described divergence in monetary policy regimes in the wake of the financial crisis should be expected to affect forex spillovers. We test this by regressing the rolling sample spillover measure on a set of shadow short rates for major countries/currency blocks (e.g., the US, EU, Japan, and the UK) and relevant control variables.\footnote{We use oil prices to control for developments in commodity prices and the S\&P500 index to control for developments in financial markets.} The shadow short rate is well suited to such an analysis because it can capture the monetary policy stance even when quantitative easing effectively pushes the short end of the yield curve into negative territory. 
\begin{table}[ht]
\scriptsize
\caption{The table shows estimation results from regressing total connectedness on major shadow short rates, namely United States (US), European Union (EU), Japan, and United Kingdom (UK). *** denotes p-value \textless 0.001,  ** p-value \textless 0.01, and  * p-value \textless 0.05.}
\centering
\begin{tabular}{lllll}
\toprule
            & US       & EU     &    Japan     & UK        \\
\cmidrule{2-5}
(Intercept) & 67.30$^{***}$ & 67.30$^{***}$ & 67.30$^{***}$ & 67.30$^{***}$ \\
            & (0.06)    & (0.07)    & (0.07)    & (0.05)    \\
Oil         & -0.62$^{***}$ & -3.05$^{***}$ & -2.33$^{***}$ & -1.59$^{***}$ \\
            & (0.07)    & (0.10)    & (0.08)    & (0.06)    \\
SP500       & -1.29$^{***}$ & 1.60$^{***}$  & 3.01$^{***}$  & -0.64$^{***}$ \\
            & (0.06)    & (0.14)    & (0.16)    & (0.06)    \\
US       & 3.02$^{***}$  &           &           &           \\
            & (0.06)    &           &           &           \\
EU       &           & 3.50$^{***}$  &           &           \\
            &           & (0.17)    &           &           \\
Japan       &           &           & 4.53$^{***}$  &           \\
            &           &           & (0.16)    &           \\
UK       &           &           &           & 2.86$^{***}$  \\
            &           &           &           & (0.05)    \\
\cmidrule{2-5}
$R^2$          & 0.56      & 0.26      & 0.35      & 0.61     \\
\bottomrule
\label{ratesregression}
\end{tabular}
\end{table}

In Table \ref{ratesregression}, we show that, after controlling for relevant economic developments, the shadow short rates exhibit explanatory power for the spillover measure. Specifically, an increase in the shadow rate is associated with an increase in forex volatility spillovers. The effect is stronger for Japan but similar for the rest of the countries considered. Based on the results, we conclude that the increased volatility and spillovers among currencies from 2013 on are chiefly rooted in different monetary policy regimes.

Second, from Figure \ref{Fig2}, we can further gauge information on the total connectedness among the currencies and crude oil (solid bold line). By adding crude oil into the set of currencies, we create a hypothetical portfolio that reflects the gradual financialization of crude oil. A general observation is that by combining crude oil with the set of currencies, the total connectedness of a mixed portfolio is lower over the observed time period than the total connectedness of the forex portfolio. The only exceptions are 2010 and a period running into 2012, when average crude oil prices were at historically high levels. Such crude oil price development is very likely behind the increase in the total volatility spillovers between the two markets. Higher connectedness among assets means that a portfolio composed of those assets is more volatile and less stable. From a practical point of view, lower stability requires frequent portfolio re-compositions. Since the mixed oil and forex portfolio exhibits lower connectedness in general, it represents the more stable investment option.

We can further enrich our observations from Figure \ref{Fig2} by quantifying directional spillovers among the analyzed assets. We compute directional spillovers and show how volatility from a specific asset (oil or currency) transmits to other assets in the portfolio (``contribution TO"). Similarly, we are also able to show the opposite link of the extent of spillovers going from a group of assets to a specific asset (``contribution FROM").

We assess directional spillovers on the forex market first. In Table \ref{spill1}, we present the aggregate effect of how specific currencies transmit and receive spillovers or, in other words, how the shocks to one currency impact other currencies (a dynamic presentation is available in the supplementary material). The highest values lie on a diagonal and represent the extent to which the own volatility of a specific currency affects its own subsequent volatility. Other values in the matrix show the volatility spillover impact between currency pairs. An interesting and intuitive observation is that shocks to each of the two commodity currencies (AUD and CAD) impact these currencies to larger extent than they do the rest of the currencies. Similarly, the euro and British pound have substantial volatility spillovers between one another. Finally, the Swiss franc and Japanese yen seem to be the calmest currencies in the portfolio, a finding that indirectly supports their status as safe havens.

\begin{table}[ht]
 \scriptsize
\caption{Volatility connectedness: currencies only}
\centering
\begin{tabular}{rcccccc|c}
\toprule
       &    AUD &    GBP &    CAD &    EUR &    JPY &    CHF &   FROM \\ 
       \cmidrule{2-8}
  AUD  &  32.24 & 15.18 & 16.92 & 13.98 & 10.21 & 11.47 & 11.29 \\
  GBP  &  15.92 & 33.13 & 13.66 & 15.86 &  9.94 & 11.48 & 11.14 \\
  CAD  &  20.81 & 15.88 & 30.91 & 12.38 &  8.70 & 11.33 & 11.52 \\
  EUR  &  16.22 & 16.22 & 11.25 & 28.12 &  8.41 & 19.79 & 11.98 \\
  JPY  &  15.72 & 15.47 & 10.54 & 12.63 & 33.18 & 12.46 & 11.14 \\
  CHF  &  15.18 & 13.08 & 11.63 & 22.19 &  9.62 & 28.29 & 11.95  \\
   \cmidrule{2-8}
  TO   &  13.98 & 12.64 & 10.66 & 12.84 &  7.81 & 11.09 & 			\\
       & 		& 		 &			& 		& 		& 		&  69.02 \\
       \bottomrule
\label{spill1}
\end{tabular}
\end{table}

\begin{table}[ht]
 \scriptsize
\caption{Volatility connectedness: currencies and crude oil}
\centering
\begin{tabular}{rccccccc|c}
\toprule
             &    AUD &    GBP &    CAD &    EUR &    JPY &    CHF & Crude Oil &  FROM \\ 
              \cmidrule{2-9}
  AUD        &       30.05 & 13.49 & 15.09 & 12.42 &  9.56 & 10.13 &  9.26 &  9.99  \\
  GBP        &       14.37 & 31.10 & 12.09 & 14.31 &  9.37 & 10.28 &  8.47 &  9.84 \\
  CAD        &       18.64 & 13.84 & 28.44 & 10.67 &  8.05 &  9.79 & 10.57 & 10.22 \\ 
  EUR        &       14.75 & 14.69 &  9.90 & 26.38 &  8.00 & 18.37 &  7.92 & 10.52 \\
  JPY        &       14.59 & 14.27 &  9.62 & 11.68 & 31.85 & 11.57 &  6.42 &  9.74 \\
  CHF        &       13.83 & 11.82 & 10.31 & 20.70 &  9.24 & 26.94 &  7.16 & 10.44 \\ 
  Crude Oil  &      12.01 &  8.96 & 11.60 &  7.94 &  5.14 &  5.82 & 48.54 &  7.35  \\

  \cmidrule{2-9}
  TO         &       12.60 & 11.01 &  9.80 & 11.10 &  7.05 &  9.42 &  7.11 &   \\
              & 		& 		 &			& 		& 		& 		&    &  68.10  \\
              \bottomrule
\label{spill2}
\end{tabular}
\end{table}

Next, we add crude oil to the portfolio and present the bilateral volatility impacts in Table \ref{spill2}. We observe that crude oil's own volatility dominates this asset and that pattern of volatility spillovers among the currencies remains same as that observed in Table \ref{spill1}. However, a new observation is that shocks to crude oil transfer to currencies to a lesser extent than shocks to currencies spill over to crude oil. 

In terms of specific observations, volatility spillovers among crude oil, Japanese yen and Swiss franc appear to be quite balanced in both directions between the oil and forex markets. The two safe haven currencies seem to be resistant to the shocks to crude oil but, at the same time, their volatilities rarely transmit to oil. Further, shocks from the two commodity currencies (AUD and CAD) affect crude oil to greater extent than those from the rest of the currencies. We conjecture that, despite being partially a target of financial speculators, crude oil is primarily a commodity, and its link to other commodities that Australia and Canada export might drive this result. Our conjecture is in line with the evidence of \cite{kohlscheen2017walk}, who show that variation in commodity prices has an effect on the nominal exchange rates of commodity-exporting countries at high frequency that goes beyond the impact of global risk appetites. With respect to the two commodity currencies in question,  \cite[p. 131]{kohlscheen2017walk} show that commodity price variation alone explains more than 23 percent of the variation in the USD exchange rate of the Australian and Canadian currencies. Similarly, \cite{ferraro2015can} show the link between oil prices and the exchange rates of the AUD, CAD, and Norwegian krone.
\subsection{Asymmetric connectedness}
\label{sec:results2}
\begin{figure}
   \centering
   \includegraphics[width=4in]{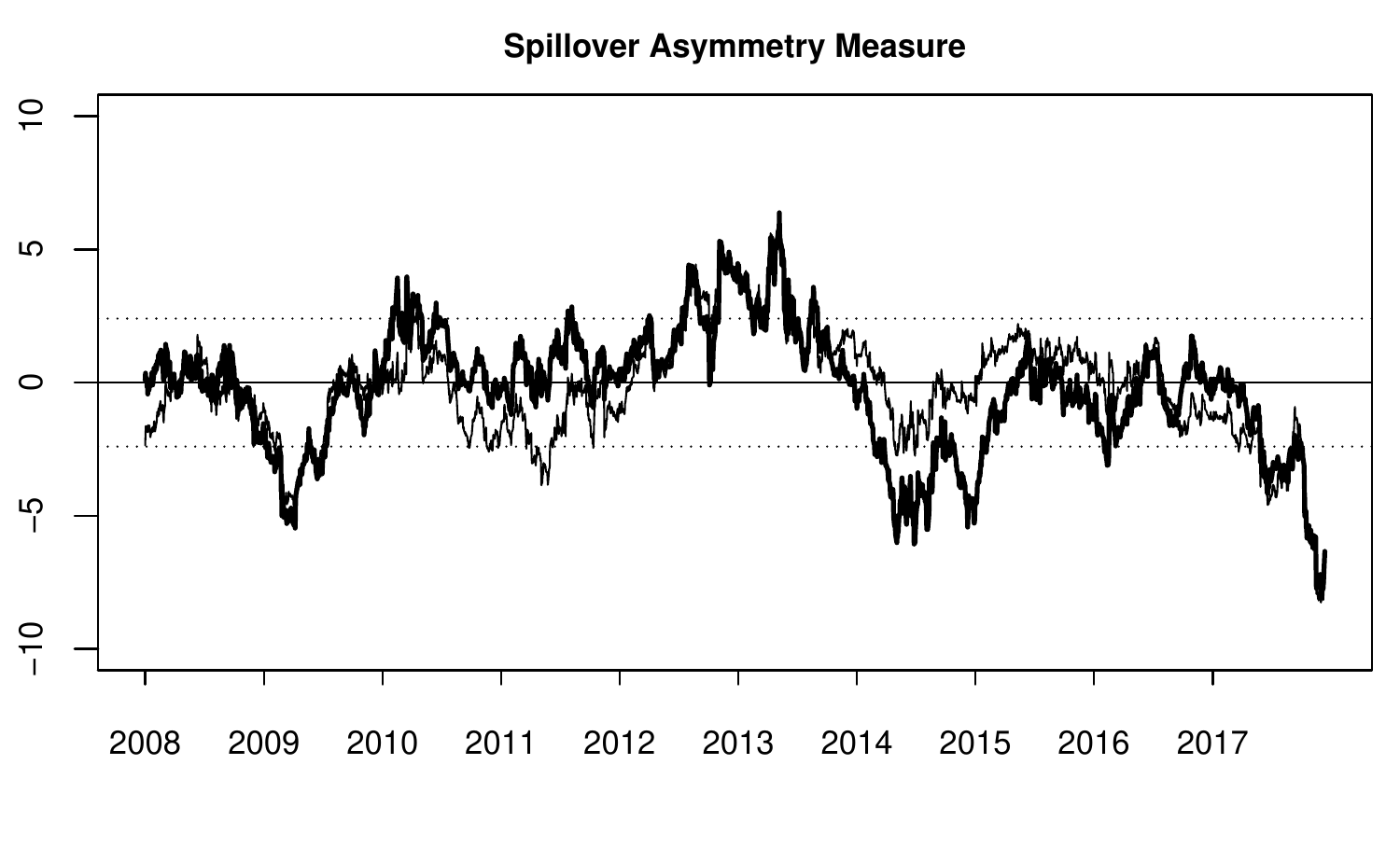}
   \caption{Spillover asymmetry measure (SAM). The solid line represents the SAM for the forex market only, while the bold solid line represents the SAM for the crude oil and forex markets. Dotted lines represent 95\% bootstrapped confidence bands.}
   \label{Fig3}
\end{figure}

We account for asymmetries in volatility shocks, and in Figure \ref{Fig3}, we plot the dynamics of the spillover asymmetry measure ($\mathcal{SAM}$) computed separately for the forex market (solid line) and the forex and oil markets (solid bold line).

The solid line being in a positive domain means that asymmetries due to positive shocks dominate asymmetries due to negative shocks. On the other hand, asymmetries due to negative shocks dominate when the solid line is situated in the negative domain. The existence of the asymmetries presented in Figure \ref{Fig3} confirms that shocks exhibit asymmetric effects on volatility spillovers between the oil and forex markets.

A general observation from Figure \ref{Fig3} is that inclusion of crude oil in the forex portfolio tends to increase the dominance of the asymmetries due to positive spillovers. The pattern is most clearly visible over the period from 2010 onwards. There are two interesting periods, however. The first one relates to the global financial crisis and the economic recovery beginning in 2009. During the first period (2009 -- 2012) the negative spillovers among the foreign currencies are substantial. The inclusion of oil into a hypothetical portfolio lowers degree of asymmetries, and the remaining asymmetries are found chiefly in the positive domain. In this respect, it is important to know that volatility spillovers among oil and oil-based commodities substantially changed after the global financial crisis: the magnitude of spillovers increased, but their asymmetries declined (see \cite{barunik2015} for details).\footnote{\cite{barunik2015} quantify asymmetries in the volatility spillovers of petroleum commodities: crude oil, gasoline, and heating oil. They show that the increase in volatility spillovers after 2001 correlates with the progressive financialization of the commodities. After 2008 (the financial crisis and advent of tight oil production), asymmetries in total and directional spillovers markedly decline. Volatility spillovers exhibit asymmetries, and spillovers due to negative (price) returns occur more often and are larger than positive ones. In terms of the directional transmission of spillovers, no petroleum commodity dominates other commodities.} Two factors seem to be behind this pattern. One is the progressive financialization of oil-based commodities that occurred after 2000 \citep{bunn2017fundamental}.\footnote{It is fair to note that the effect of oil's financialization may be overstated. In connection with the role of financial speculators, \cite[p. 465]{kilianMurphy2014role} argue that there is ``no systematic upward movement in the real price of oil after 2003 associated with speculative demand shocks." Further, \cite{kilianLee2014quantifying} present evidence undermining the hypothesis that the financialization of oil markets caused oil price increases after 2003.} The second factor is the beginning of substantial tight oil exploration from very low permeability shale, sandstone, and carbonate formations and an increase in U.S. oil production, later resulting in a supply shock in global markets. We conjecture that lower asymmetries related to oil (and oil commodities) documented by \cite{barunik2015} produce this beneficial result.

The second period exhibits an entirely different pattern. A non-negligible increase in the extent of negative spillovers is visible around 2014. The pattern is detected for both forex and forex-oil mixed portfolios. An important feature is that the inclusion of the crude oil correlates with a further increase in negative spillovers. This excess is underlined by an intuitive explanation, however. Oil prices dropped sharply in 2014, and the drop was so large that it augmented asymmetry in spillovers of the forex-oil portfolio. However, it has to be noted that the extent of the asymmetry is not dramatic: asymmetries in spillovers in either the forex portfolio or forex-oil mixed portfolio are relatively low when compared to other assets as shown in \cite{barunik2016asymmetric}. There is a good reason for this pattern: increasing financialization of oil has led to increases in portfolio sizes and the number of transactions; such increases in trading activity might induce a decline in spillover asymmetries via the price-setting mechanism on the market \citep{barunik2015}. A similar logic applies to the forex trading in major currencies that has recorded increasing volumes of transactions over time \citep{bis2016}. Thus, the combined effect of the massive volume of transactions involving the two types of assets (oil and currencies) contributes to lower asymmetries in volatility spillovers. The temporary increase in negative spillovers due to the sharp drop in oil prices represents an isolated factor that is likely behind the asymmetry increase in 2014. The dynamics of directional spillovers are available in the supplementary material.

Finally, \cite{aloui2013conditional} argue that higher oil reserves and better production positions of a country help to reduce the extreme dependence of its foreign exchange market on oil price fluctuations. This statement indirectly implies the existence of lower volatility spillovers between oil and forex markets when oil reserves are solid and production is stable. In a similar way, high connectedness between oil and forex markets should prevail during periods of unavailability. Low inventories of crude oil force refineries to buy extra crude oil and may also lead to supply problems for gasoline and other petroleum products; low inventories of crude oil then likely cause price volatility and spillovers \citep{barunik2015}. The above-described link is supported by the fact that the timing of the increased negative spillovers during 2010 -- 2011 and 2013 -- 2014 shown in Figure \ref{Fig3} match the occurrence of the negative spillovers among oil and oil-based commodities reported by \cite[Figure 3, panel b]{barunik2015}.

\subsection{Frequency connectedness}

 \begin{figure}
   \centering
   \includegraphics[width=4in]{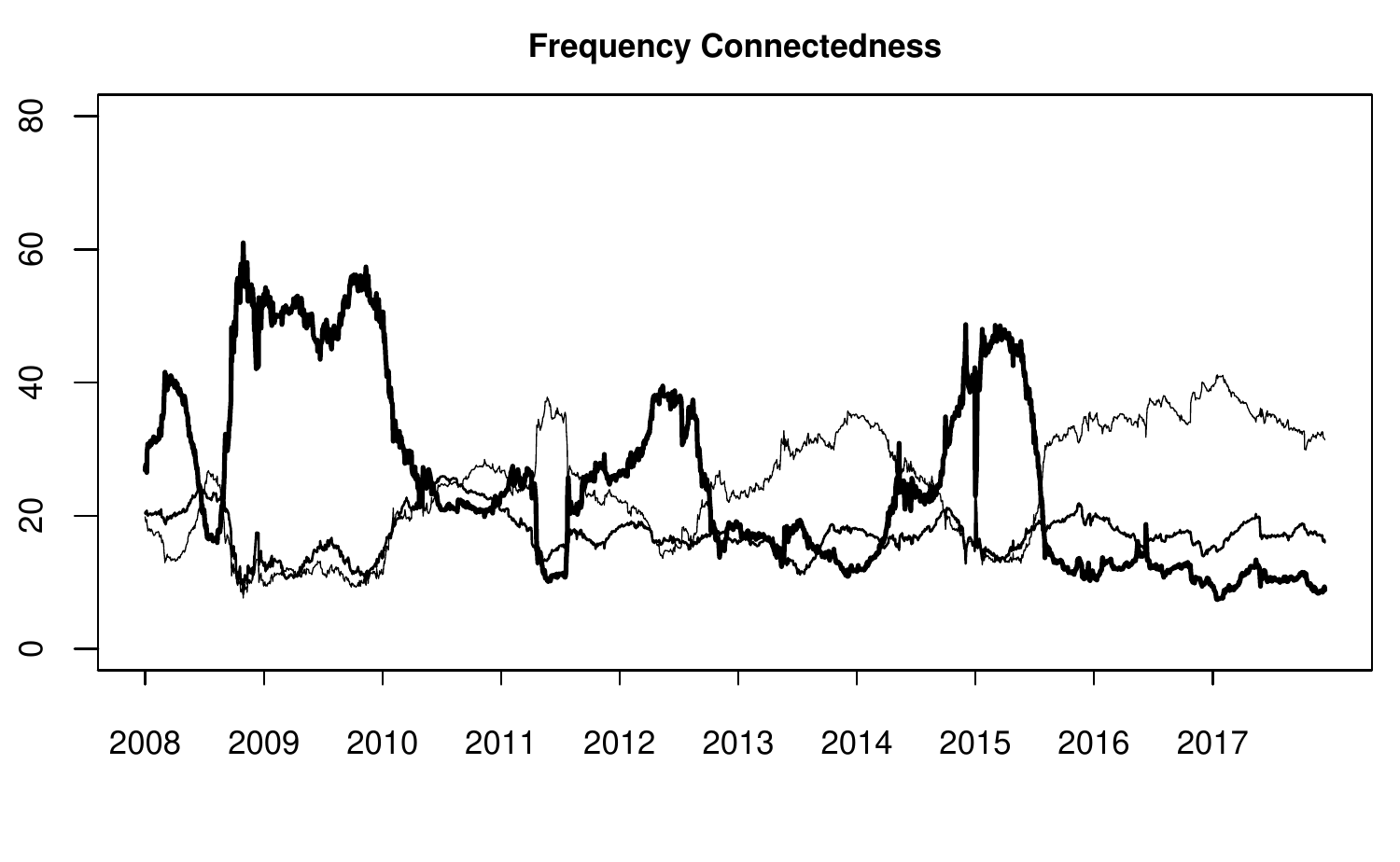}
   \caption{Dynamic frequency connectedness. The frequency connectedness at the short-term horizon defined at $d_1 \in [1,5]$ days is marked with the solid line, the medium-term horizon defined at $d_2 \in (5,20]$ days is depicted by the medium bold line, and the long-term horizon defined at $d_3 \in (20,300]$ days is represented with the bold line. Note that all lines through the frequency bands $d_s$ sum to the total connectedness.}
   \label{Fig4}
\end{figure}

In Figure \ref{Fig4}, we present plots of the frequency connectedness. The frequencies at which the frequency connectedness is computed can be understood as different investment horizons.\footnote{The short-term horizon of [1,5] days represents a business week, the medium-term horizon of (5,20] days represents a business month, and the long-term horizon of (20,300] days stands for a business year. These definitions represent different horizons from an investment perspective. Other horizons that represent reasonable alternative intervals deliver similar results. However, the plotted distinctions are less sharp than those provided by our choice of business week, month and year.} The frequency connectedness is computed for the mixed forex-oil portfolio. Three lines represent the extent of connectedness at three investment horizons. The short-term horizon is represented by a simple solid line and might be attributed to short-term investment strategies adopted, for example, in technical analysis. The bold line represents the long-run connectedness. It reflects the long-term investment horizon associated with, for example, funds or investors oriented toward developments in economic fundamentals. The medium bold line captures the medium-term horizon. 

Further, the reason that investors favor different investment horizons (represented by frequencies) comes from the formation of their preferences. \cite{ortu2013long} develop an asset pricing model in which consumption responds to shocks due to heterogeneous preference choices. Further, \cite{bandi2017horizon} argue that investors may not focus on very high-frequency components of consumption representing short-term noise but on lower frequency components of consumption growth with heterogeneous periodicities instead. As a result, various sources of connectedness might create short-, medium-, and long-term systemic risk. Thus, when studying connectedness, we should focus on linkages with various degrees of persistence underlying systemic risk, as these are likely driving the differences over different investment horizons.

In Figure \ref{Fig4}, we show that the level of the medium-term connectedness is lowest and the most stable. The dynamics of the short- and long-term connectedness are remarkably different. First, on three occasions, the long-term connectedness dramatically heightens. These increases correlate with (i) the global financial crisis from mid-2008 into early 2010, (ii) the European sovereign debt crisis in 2012, and (iii) the decline in oil prices in 2014. Second, the dynamics of the short-term connectedness represent quite a different picture. From the beginning of the global financial crisis until early 2010, the short-term connectedness (i) evolves in a very similar way as the medium-term connectedness and (ii) remains quite low when compared to the long-term connectedness. Later on, short-term connectedness always substantially rises after the long-term connectedness sharply declines. The combined effect of the short- and long-term connectedness seems to form a major part of the total connectedness, while medium-term shocks play only a minor role. The finding bears implications with respect to portfolio management and hedging strategies. The dynamics of directional frequency spillovers are available in the supplementary material.

The sharp differences between the long-term and short/medium-term connectedness should be attributed to the differences in how investors perceive the stability of the economic and financial system. To do so, we complement the above observations by an auxiliary analysis on the extent to which spillover activity is driven by liquidity shocks (i.e., due to the crash risk hypothesis) or uncertainty shocks. To that end, we regress the three frequency spillover measures on the TED spread (to measure funding liquidity) and the VIX (to measure uncertainty).

\begin{table}[ht]
 \scriptsize
\caption{The table shows estimation results from regressing short-term, medium-term, and long-term connectedness on the TED spread and VIX. *** denotes p-value \textless 0.001,  ** p-value \textless 0.01, and  * p-value \textless 0.05.}\centering
\begin{tabular}{lllllllll}
\toprule

                & \multicolumn{2}{c}{Short}    & & \multicolumn{2}{c}{Medium}   & & \multicolumn{2}{c}{Long}      \\
\cmidrule{2-3} \cmidrule{5-6} \cmidrule{8-9}
(Intercept)  & 24.63$^{***}$ & 24.63$^{***}$ && 17.36$^{***}$ & 17.36$^{***}$ && 25.25$^{***}$ & 25.25$^{***}$ \\
              & (0.17)    & (0.15)    && (0.07)    & (0.07)    && (0.26)    & (0.23)    \\
TED          & -2.90$^{***}$ &           && -0.24$^{***}$ &          & & 4.59$^{***}$  &           \\
             & (0.17)    &           && (0.07)    &           && (0.26)    &           \\
VIX         &           & -4.63$^{***}$ &&           & -0.53$^{***}$ &&           & 7.64$^{***}$  \\
              &           & (0.15)    &&           & (0.07)    &&           & (0.23)    \\
\cmidrule{2-3} \cmidrule{5-6} \cmidrule{8-9}
$R^2$            & 0.11      & 0.28      && 0.01      & 0.03      && 0.11      & 0.31      \\
\bottomrule
\label{vixtedregression}
\end{tabular}
\end{table}

In Table \ref{vixtedregression} we show that volatility spillovers between oil and forex markets are driven by both liquidity and uncertainty shocks but that the effect of uncertainty shocks is always stronger irrespective of the frequency considered. The impact of both types of shocks is associated with a sizable decrease in connectedness at the short frequency but only a slight decrease at the medium frequency. On the other hand, a substantial increase in volatility spillovers can be observed at long frequency connectedness. The results also corroborate our previous observation on the primary importance of the short- and long-term investment horizons.\footnote{Our results suggest that market uncertainty greatly impacts long-term connectedness. We can further support our results with the evidence in \cite[Figure 2.B]{fratzscher2014oil}, who show that during both periods of the financial and European sovereign debt crises (i) the negative correlation between exchange rates and oil prices strengthened and (ii) the uncertainty increased. The uncertainty is captured by the rolling standard deviation of the identified structural VIX shocks. The rolling window is 12 months. The graphical presentation of the evidence in \cite[Figure 2.B]{fratzscher2014oil} time-wise exactly matches increases in the long-term connectedness. Their sample ends in 2012.}

In terms of a particular asset type as a volatility source, our interpretation relies on the supportive results from the relevant literature \citep{barunik2017,van2016macroeconomic,fratzscher2014oil,aloui2013conditional}. We conjecture that the increase in the long-term connectedness during the global financial crisis originates in both oil and forex developments, while the 2012 increase should be credited predominantly to the forex market spillovers \citep{barunik2017}. The increase in the long-term connectedness in 2014 -- 2015 most likely relates to the drop in oil prices. However, as we show above, in all three cases, uncertainty dominates the heightened long-term connectedness. In addition, \cite{van2016macroeconomic} shows that higher macroeconomic uncertainty causes higher oil price volatility. This means that oil price volatility is typically higher during periods such as financial crises and recessions. A solid link between oil prices and the dollar exchange rate in the form of a negative correlation after 2000 \citep{fratzscher2014oil,aloui2013conditional} then serves as a basis for volatility spillovers between the two types of assets. The timing of the increases in long-term connectedness fully reflects the heightened volatility of oil prices and macroeconomic uncertainty evidenced in \cite[Figures 1 and 10, respectively]{van2016macroeconomic}.  

Finally, and as a technical remark, one should also note that a simple sum of the three lines presented in Figure \ref{Fig4} provides the total connectedness plotted earlier in Figure \ref{Fig2}. In this respect, the decomposition enabled by frequency connectedness serves to further a deeper understanding of the sources of connectedness.

\section{Conclusion \label{sec:conclusion}}
We analyze total, asymmetric and frequency connectedness on the oil and forex markets using high-frequency, intra-day data over the period 2007 -- 2017. We show that high forex connectedness decreases when both forex and oil markets are assessed jointly. The increased volatility and spillovers among currencies from 2013 onward are chiefly rooted in different monetary policy regimes. Since oil market development seems to be detached from monetary policy regimes, we might argue that a mixed oil and forex portfolio, with its lower connectedness in general, represents the more stable investment option. This is true despite that shocks to oil transfer to currencies to a lesser extent than vice versa. A practical implication emerges that total connectedness of the mixed forex and oil portfolio might be reduced by selecting currencies that transfer their volatility to oil to the smallest extent.

Asymmetries in forex volatility connectedness are dominated by negative shocks in general. Connectedness between both oil and forex markets is dominated by positive shocks, however. A direct implication is that adding oil to form a mixed oil and forex portfolio has the potential to alter the asymmetry in the connectedness between the two classes of assets. Asymmetries in connectedness are also relatively small. The massive volume of transactions
involving both types of assets represents enormous information flows on both markets. We conjecture that ample information is likely behind the limited extent of asymmetries in volatility spillovers.  

The dynamics of frequency connectedness differ dramatically across various investment horizons and should be credited to differences in investment preference formation. Frequency connectedness is to an extent driven by both liquidity and uncertainty shocks, and uncertainty shocks always exert a stronger impact irrespective of frequency. Both types of shocks suppress short-term connectedness but substantially raise long-term connectedness. This finding helps to explain frequency connectedness behavior, in that long-term connectedness reflects substantive features of economic development and investors' concerns. Hence, frequency connectedness can also serve as a sensitive indicator of the horizon at which markets feel uncertainty the most.

{\footnotesize{
\setlength{\bibsep}{3pt}
\bibliographystyle{chicago}
\bibliography{spillovers}
}}

\newpage
\appendix
\section*{Supplementary Material for ``Total, asymmetric and frequency connectedness between oil and forex markets''}

\section{Realized variance and semivariance \label{sec:semi}}
In this Section we briefly introduce realized measures that we use for volatility connectedness estimation. We begin with realized variance and then we describe realized semivariances. Realized measures are defined on a continuous-time stochastic process of log-prices, $p_t$, evolving over a time horizon $[0\le t \le T]$. The process consists of a continuous component and a pure jump component,
\begin{equation}
p_t=\int_0^t\mu_s ds + \int_0^t\sigma_s d W_s + J_t,
\end{equation}
where $\mu$ denotes a locally bounded predictable drift process, $\sigma$ is a strictly positive volatility process, and $J_t$ is the jump part, and all is adapted to some common filtration $\mathcal{F}$. The quadratic variation of the log prices $p_t$ is:
\begin{equation}
\label{eq:qv}
[p_t,p_t] = \int_0^t\sigma_s^2 ds+\sum_{0<s\le t}(\Delta p_s)^2,
\end{equation}
where $\Delta p_s = p_s - p_{s-}$ are jumps, if present. The first component of Eq. (\ref{eq:qv}) is integrated variance, whereas the second term denotes jump variation. \cite{andersen1998answering} proposed estimating quadratic variation as the sum of squared returns and coined the name ``realized variance" ($RV$). The estimator is consistent under the assumption of zero noise contamination in the price process.

Let us denote the intraday returns $r_k=p_k-p_{k-1}$, defined as a difference between intraday equally spaced log prices $p_0,\ldots,p_n$ over the interval $[0,t]$, then
\begin{equation}
RV=\sum_{k=1}^n r_k^2
\end{equation}
converges in probability to $[p_t,p_t]$ with $n\rightarrow \infty$.

\cite{shephard2010measuring} decomposed the realized variance into realized semivariances ($RS$) that capture the variation due to negative ($RS^-$) or positive ($RS^+$) price movements (e.g., bad and good volatility). The realized semivariances are defined as:
\begin{eqnarray}
RS^-&=& \sum_{k=1}^n \mathbbm{I}(r_k<0) r_k^2, \\
RS^+&=& \sum_{k=1}^n \mathbbm{I}(r_k\ge0) r_k^2.
\end{eqnarray}
Realized semivariance provides a complete decomposition of the realized variance, hence:
 \begin{equation}
RV=RS^- + RS^+.
\end{equation}
The limiting behavior of realized semivariance converges to $1/2\int_0^t\sigma_s^2 ds$ plus the sum of the jumps due to negative and positive returns \citep{shephard2010measuring}. The negative and positive semivariance can serve as a measure of downside and upside risk as it provides information about variation associated with movements in the tails of the underlying variable.

\clearpage
\section{Supplementary Tables and Figures}
\subsection{Directional Connectedness}

 \begin{figure}[ht]
   \centering
   \includegraphics[width=6in]{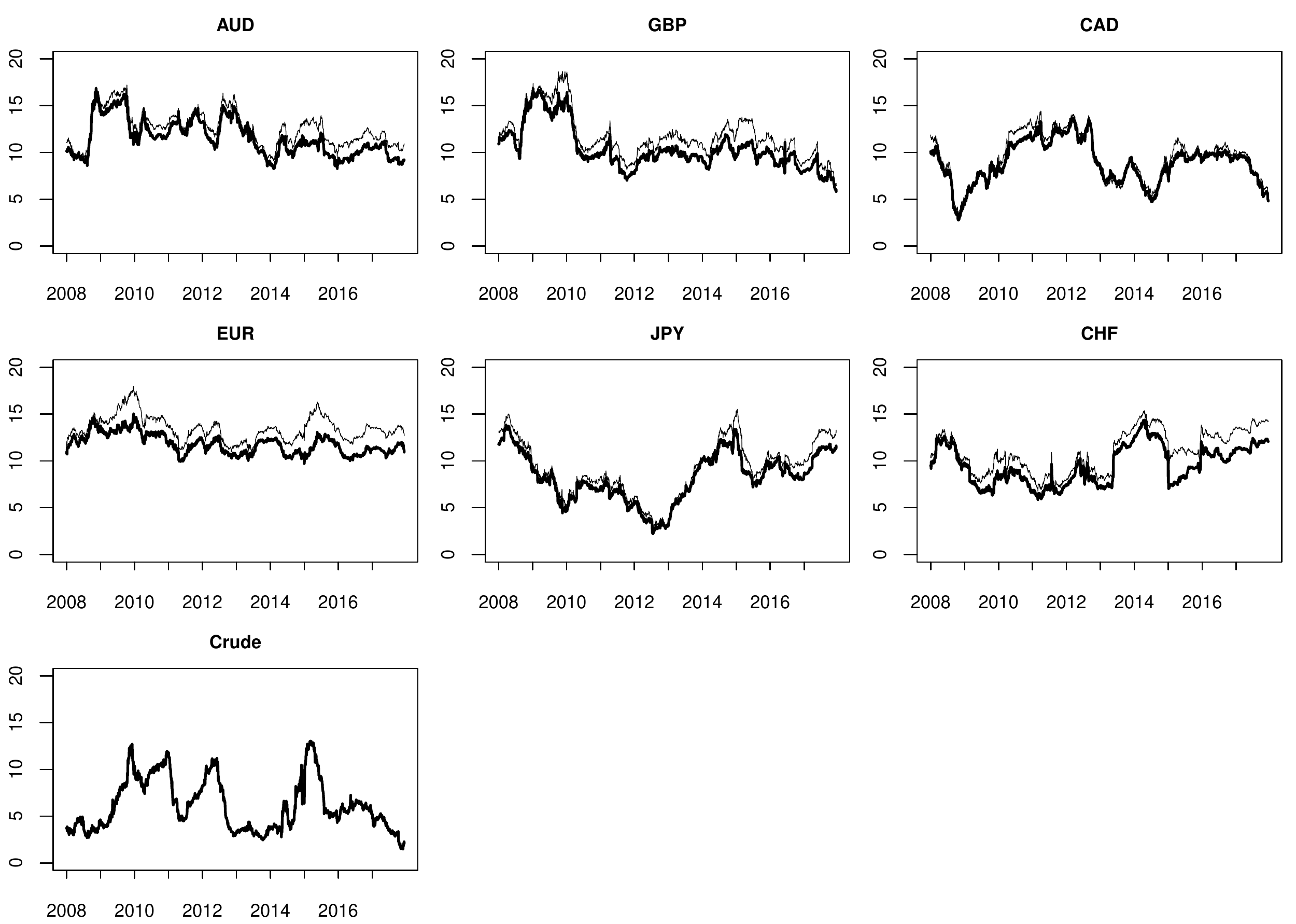}
   \caption{The directional (TO) volatility connectedness of six currencies (solid line), the directional volatility connectedness of six currencies and crude oil (bold solid line). The directional (TO) volatility connectedness quantifies how volatility from a specific asset (oil or currency) transmits to other assets in portfolio (``contribution TO'').}
   \label{app:FigTO}
\end{figure}

 \begin{figure}
   \centering
   \includegraphics[width=6in]{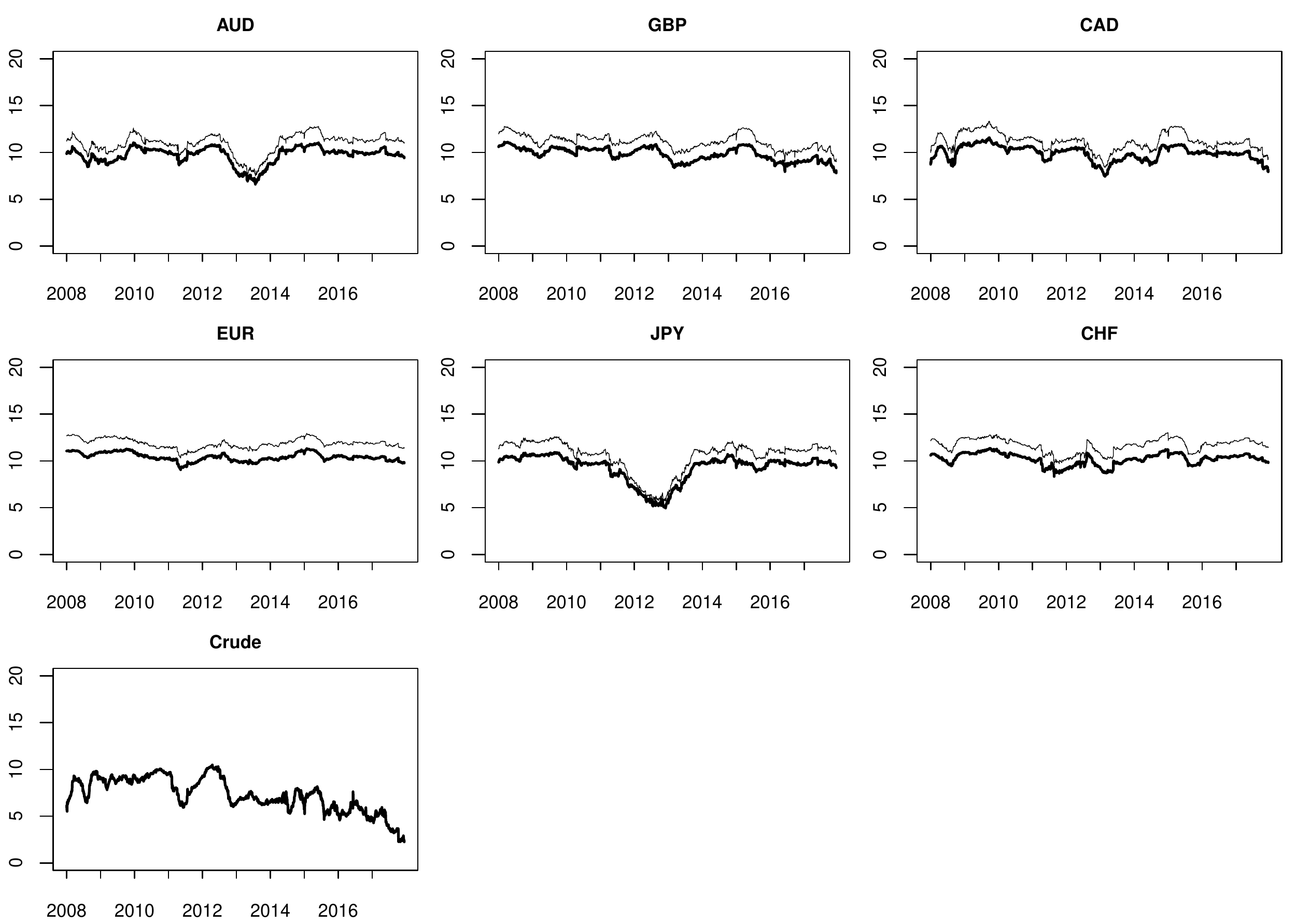}
   \caption{The directional (FROM) volatility connectedness of six currencies (solid line), the directional volatility connectedness of six currencies and crude oil (bold solid line). The directional (FROM) volatility connectedness quantifies how volatility from a group of assets  transmits to a specific asset (oil or currency) (``contribution FROM'').}
   \label{app:FigFROM}
\end{figure}

 \begin{figure}
   \centering
   \includegraphics[width=6in]{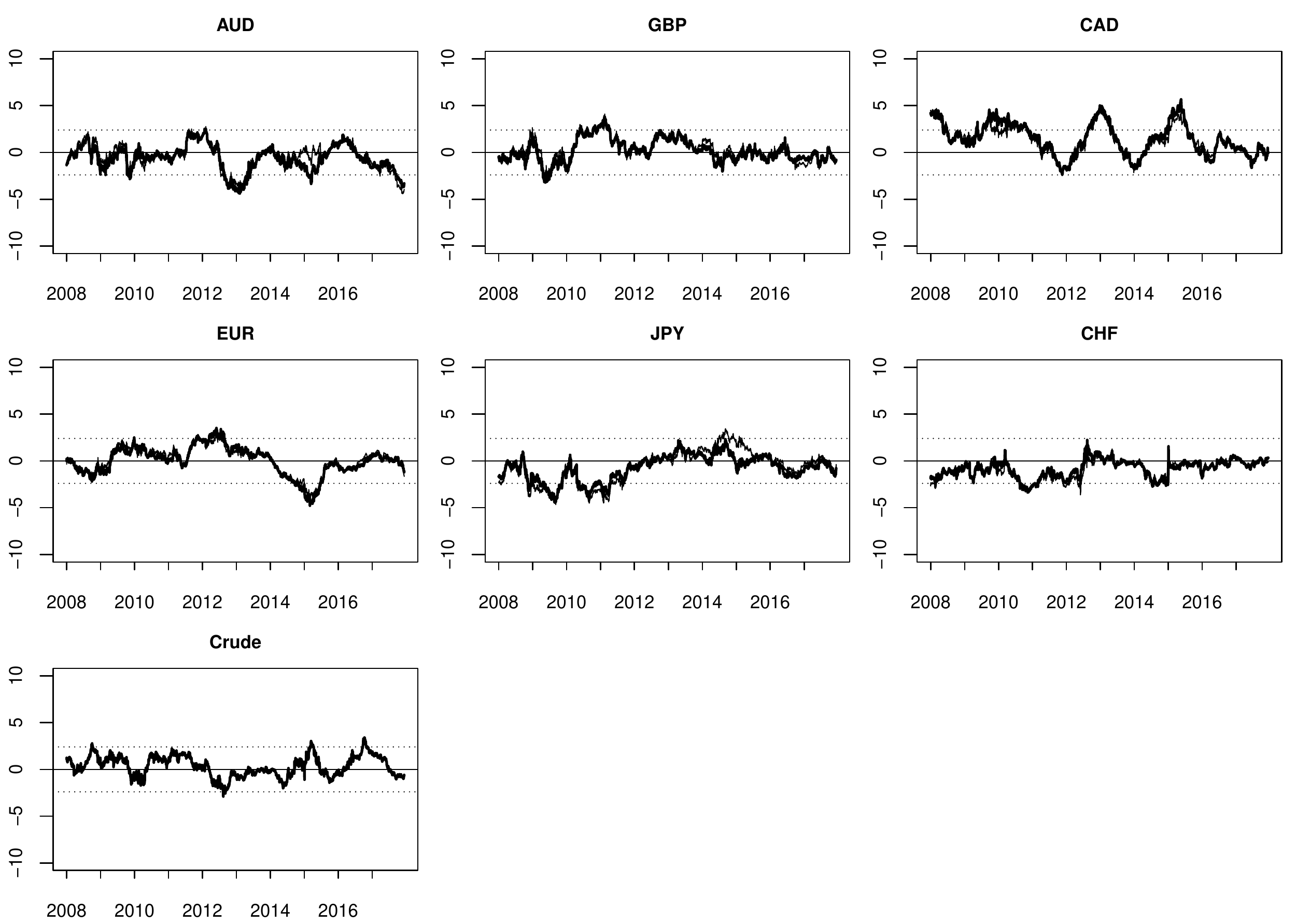}
   \caption{Directional (TO) Spillover asymmetry measure (SAM). Solid line represents the directional SAM for the forex market only, while the bold solid line represents the directional SAM for the crude oil and forex markets. 95\% bootstrapped confidence bands are shown by dotted lines. The directional SAM (TO) quantifies how asymmetry in volatility from a specific asset (oil or currency) transmits to other assets in portfolio (``contribution TO''}
   \label{app:FigTOasym}
\end{figure}

 \begin{figure}
   \centering
   \includegraphics[width=6in]{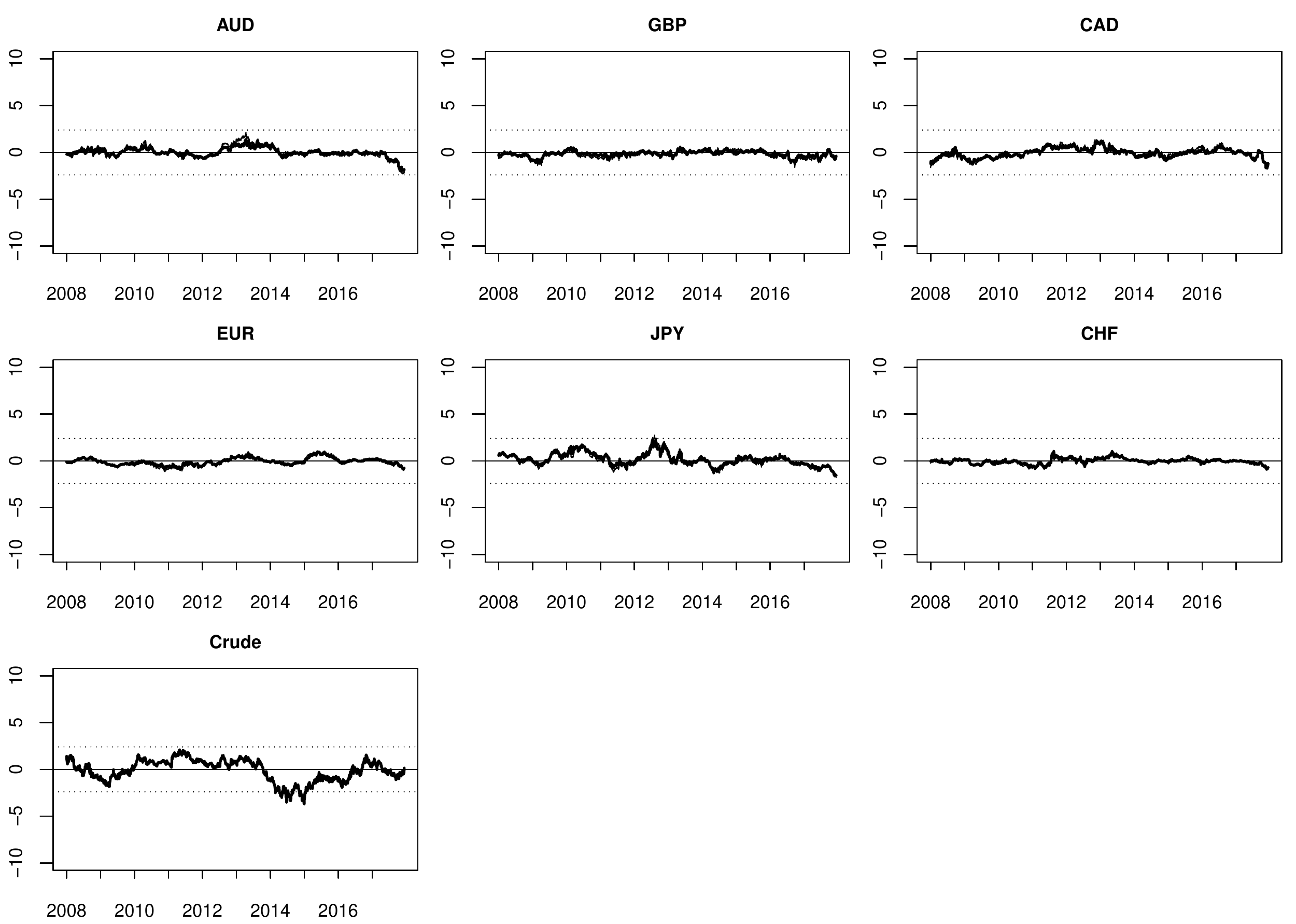}
   \caption{Directional (FROM) Spillover asymmetry measure (SAM). Solid line represents the directional SAM for the forex market only, while the bold solid line represents the directional SAM for the crude oil and forex markets. 95\% bootstrapped confidence bands are shown by dotted lines. The directional SAM (FROM) quantifies how asymmetry in volatility from a group of assets  transmits to a specific asset (oil or currency) (``contribution FROM'').}
   \label{app:FigFROMasym}
\end{figure}

 \begin{figure}
   \centering
   \includegraphics[width=6in]{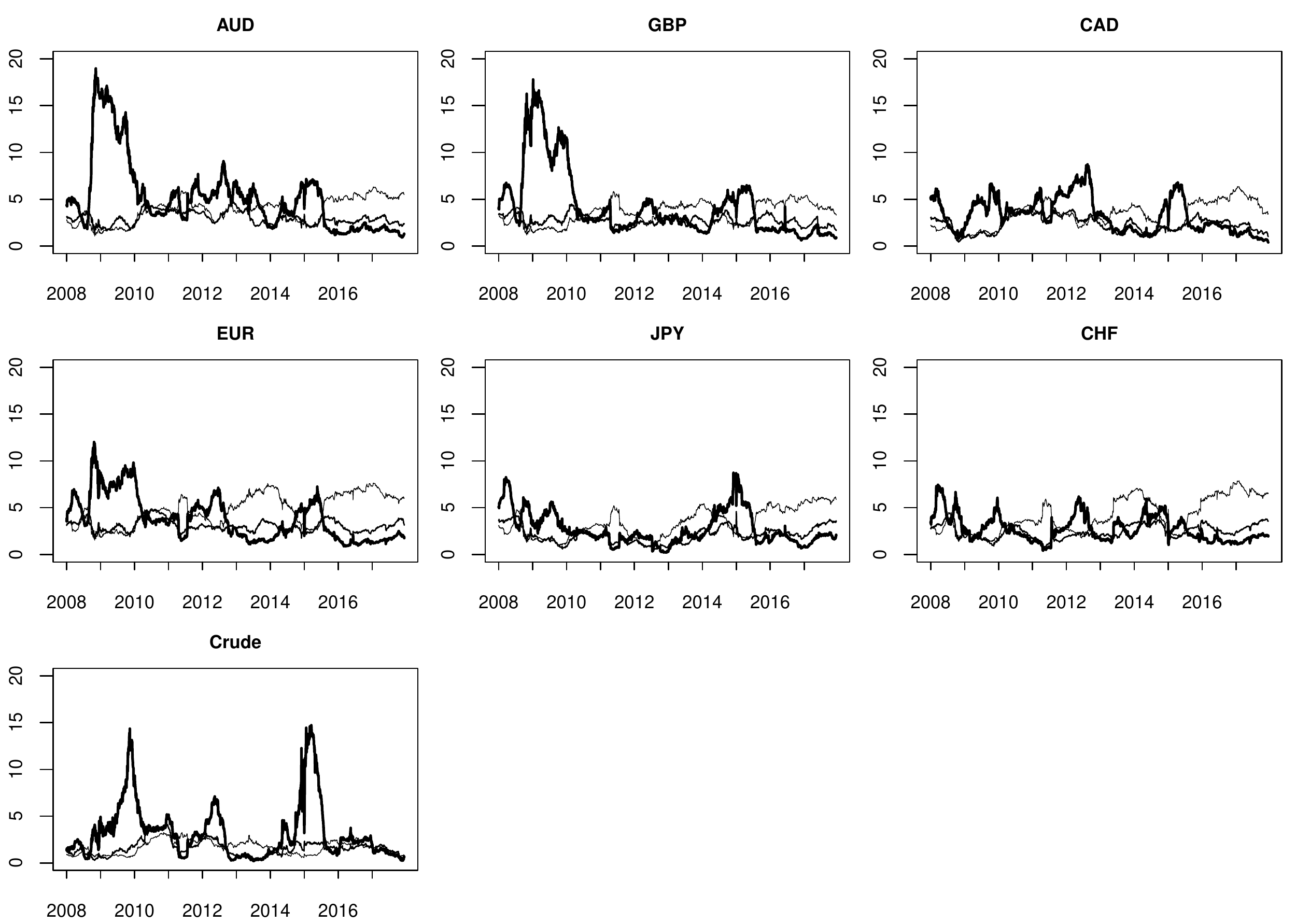}
      \caption{Directional (TO) frequency connectedness. The frequency connectedness at short-term horizon defined at $d_1 \in [1,5]$ days in solid line, medium-term horizon defined at $d_2 \in (5,20]$ days medium bold line, and long-term horizon defined at $d_3 \in (20,300]$ days in bold line. Note that all lines through the frequency bands $d_s$ sum to the total connectedness. The directional (TO) frequency connectedness quantifies how volatility (measured at given frequency) from a specific asset (oil or currency) transmits to other assets in portfolio (``contribution TO'').}
   \label{app:FigTOfreq}
\end{figure}

 \begin{figure}
   \centering
   \includegraphics[width=6in]{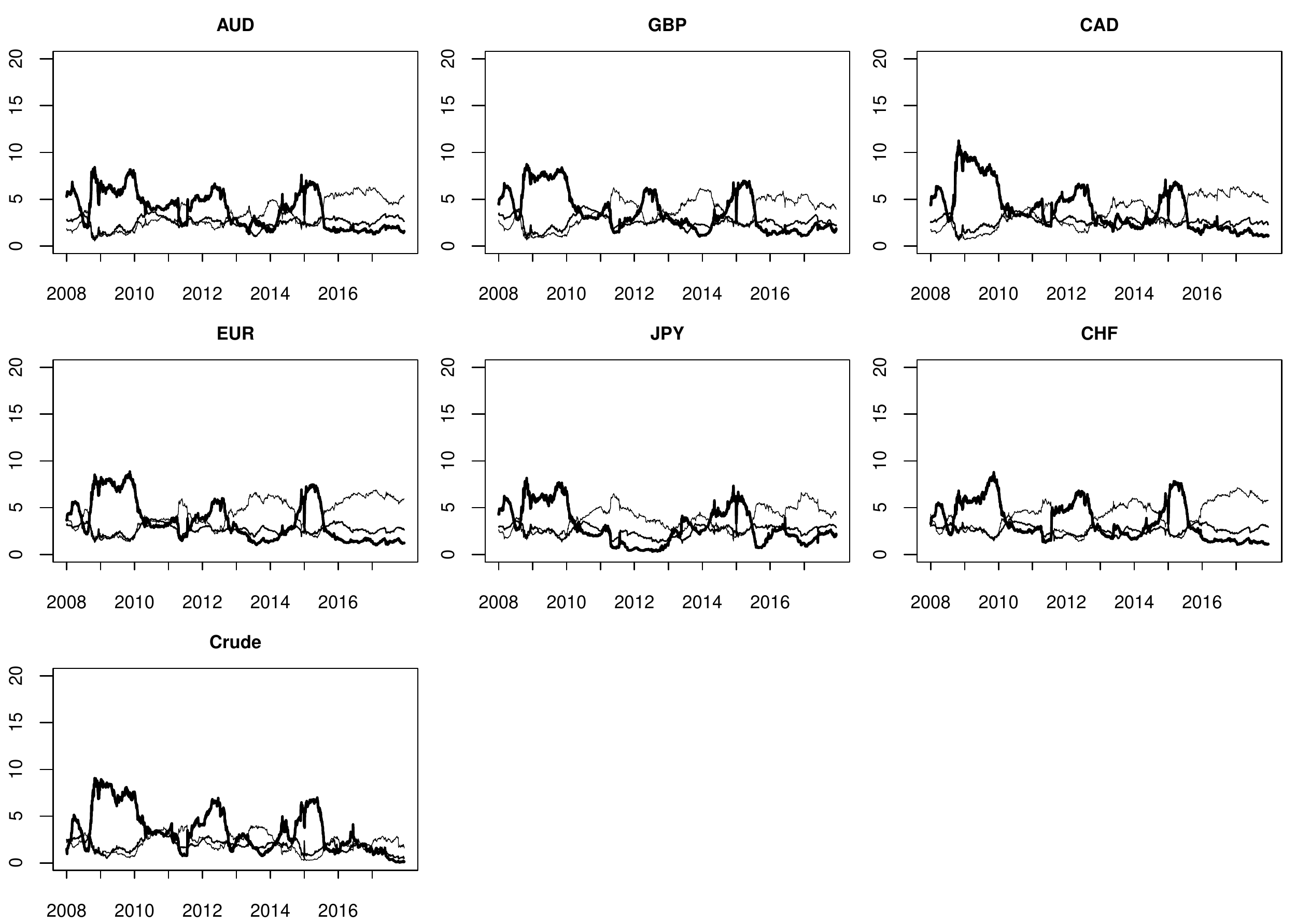}
      \caption{Directional (FROM) frequency connectedness. The frequency connectedness at short-term horizon defined at $d_1 \in [1,5]$ days in solid line, medium-term horizon defined at $d_2 \in (5,20]$ days medium bold line, and long-term horizon defined at $d_3 \in (20,300]$ days in bold line. Note that all lines through the frequency bands $d_s$ sum to the total connectedness. The directional (FROM) frequency connectedness  quantifies how volatility (measured at given frequency) from a group of assets  transmits to a specific asset (oil or currency) (``contribution FROM'').}
   \label{app:FigFROMfreq}
\end{figure}

\clearpage
\subsection{Sensitivity Analysis}
 \begin{figure}[h!]
   \centering
   \includegraphics[width=4in]{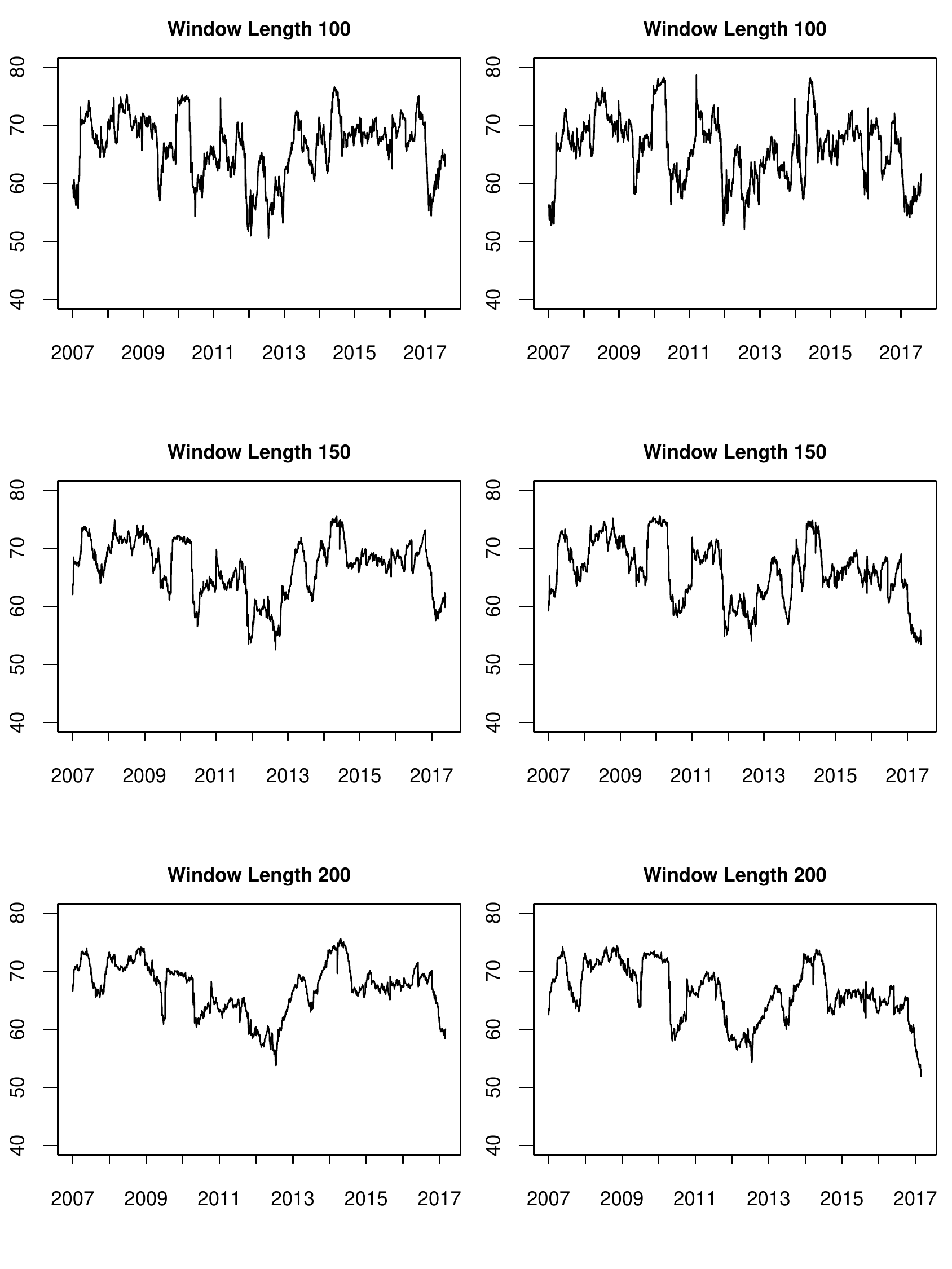}
   \caption{Sensitivy Analysis: Total volatility connectedness of six currencies (left column), and of six currencies and crude oil (right column). Total connectedness computed for different window lengths (top row), horizons (middle row), and VAR lengths (bottom row).\textbf{}}
   \label{app:Figrobust1}
\end{figure}

 \begin{figure}
   \centering
   \includegraphics[width=4in]{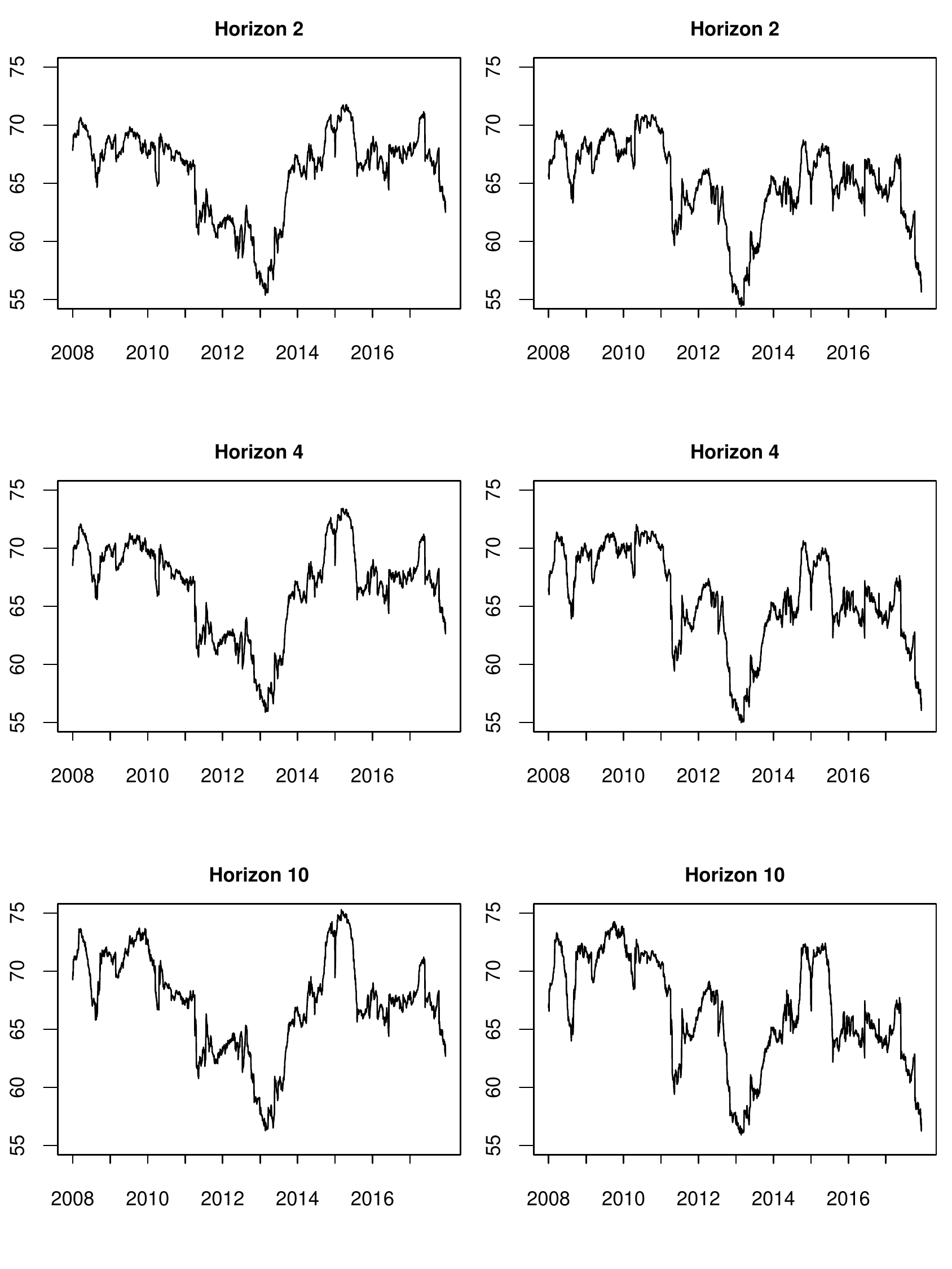}
   \caption{Sensitivy Analysis: Total volatility connectedness of six currencies (left column), and of six currencies and crude oil (right column). Total connectedness computed for different window lengths (top row), horizons (middle row), and VAR lengths (bottom row).\textbf{}}
   \label{app:Figrobust2}
\end{figure}

 \begin{figure}
   \centering
   \includegraphics[width=4in]{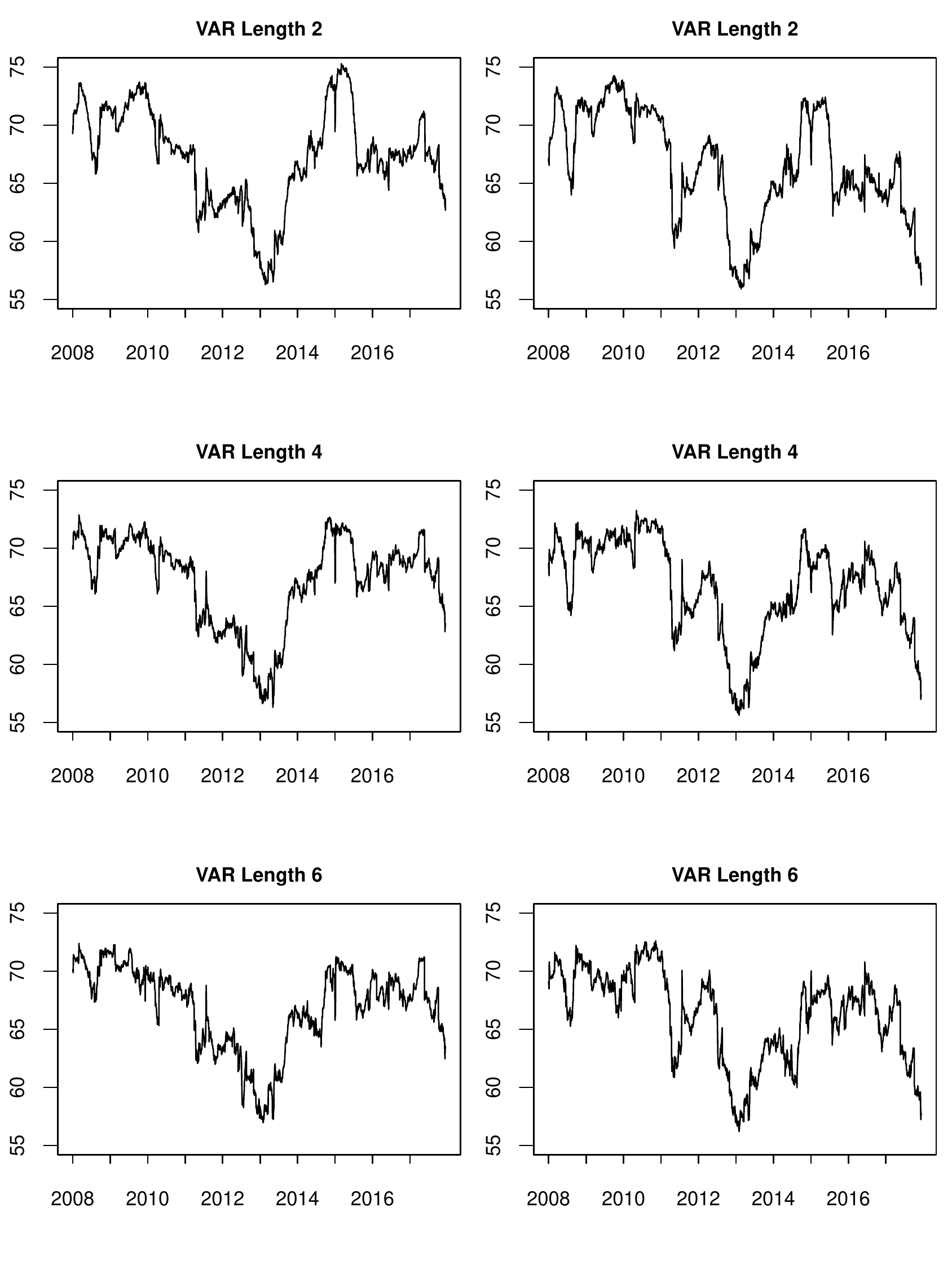}
   \caption{Sensitivy Analysis: Total volatility connectedness of six currencies (left column), and of six currencies and crude oil (right column). Total connectedness computed for different window lengths (top row), horizons (middle row), and VAR lengths (bottom row).\textbf{}}
   \label{app:Figrobust3}
\end{figure}

\end{document}